\documentclass[pdflatex,sn-mathphys-num]{sn-jnl}

\usepackage{amsmath,amssymb,amsfonts,bm}
\usepackage{needspace}
\usepackage{float}
\usepackage{tikz}
\usetikzlibrary{arrows.meta,calc}

\hypersetup{
  pdftitle={Exact Matching-Polynomial Solution of the Periodic Baxter-Fendley ZN Clock Chain},
  pdfauthor={Yuguan Li, D. C. Liu, and Murray T. Batchelor},
  pdfsubject={Exact matching-polynomial solution of the periodic ZN clock chain},
  pdfkeywords={periodic clock chain, matching polynomial, cyclic Weyl algebra},
  pdfcreator={LaTeX with the Springer Nature sn-jnl class}
}

\theoremstyle{thmstyleone}
\newtheorem{proposition}{Proposition}[section]
\newtheorem{lemma}{Lemma}[section]
\newtheorem{corollary}{Corollary}[section]
\theoremstyle{thmstylethree}
\newtheorem{definition}{Definition}[section]
\newtheorem{remark}{Remark}[section]
\newtheorem{assumption}{Assumption}[section]
\newtheorem{algorithm}{Algorithm}[section]
\theoremstyle{thmstyleone}
\newtheorem{theorem}[definition]{Theorem}

\begin{document}

\title[Exact Matching-Polynomial Solution of the Periodic \(Z_N\) Clock Chain]
{Exact Matching-Polynomial Solution of the Periodic Baxter-Fendley
  \texorpdfstring{\(Z_N\)}{ZN} Clock Chain}

\author[1]{\fnm{Yuguan} \sur{Li}}
\email{yuguan.li@anu.edu.au}

\author[1]{\fnm{D. C.} \sur{Liu}}
\email{dongchang.liu@anu.edu.au}

\author*[1]{\fnm{Murray T.} \sur{Batchelor}}
\email{murray.batchelor@anu.edu.au}

\affil*[1]{\orgdiv{Mathematical Sciences Institute},
  \orgname{Australian National University},
  \orgaddress{\city{Canberra}, \postcode{2601}, \state{ACT},
  \country{Australia}}}

\abstract{
The periodic-boundary non-Hermitian Baxter-Fendley \(Z_N\) chain does not have the same free-parafermionic structure as the open-boundary chain. We construct a matching polynomial for the corresponding cycle of its cyclic Weyl algebra. Its coefficients form conserved quantities containing the periodic Hamiltonian, while the matching polynomial itself realizes a Yang-Baxter transfer matrix. Root-of-unity fusion gives a second matching-polynomial closure and a system of spectral equations in each charge sector. The normalized energy resultant of these equations coincides with the charge-sector characteristic polynomial, yielding the complete finite-size energy spectrum with algebraic multiplicity. For positive reflection-symmetric chains, we prove that the periodic-boundary ground state is the unique analytic continuation of the open-boundary ground state. This yields an efficient large-size ground-state algorithm without enumerating the full spectrum. 
For the homogeneous three-state chain, seam-response asymptotics and finite-size ground-state
calculations support two reciprocal candidate critical couplings distinct from the self-dual point, 
while the full periodic phase diagram is left for future work.
}

\maketitle

\section{Introduction}\label{sec:introduction}

The open Baxter-Fendley \(Z_N\) clock chain is exactly solvable in terms of
independent free parafermions, although periodic closure destroys the
corresponding description in terms of independent quasienergies~\cite{AlcarazBatchelor2018}.  With open
boundary conditions (OBC), the complete spectrum is assembled from \(L\)
independent quasienergies, each with \(N\) possible occupations
\cite{Baxter1989,Fendley2014,BatchelorHenryLu2023}.  Under periodic boundary
conditions (PBC), the Weyl-algebra graph closes a directed path into a directed
cycle, while the Fradkin-Kadanoff closing bond becomes a nonlocal
string~\cite{Fendley2014,MannElmanWoodChapman2025}.  These features obstruct a
direct extension of the independent-quasienergy construction to the PBC.  For
\(N>2\), direct investigations of this periodic Hamiltonian have used numerical
diagonalization and finite-size extrapolation~\cite{AlcarazBatchelor2018}. 
To the best of our knowledge, no explicit finite-size solution specialized 
to the periodic Baxter-Fendley chain 
has been given~\cite{Fendley2014,AlcarazBatchelor2018,Alcaraz2026Nonhomogeneous}.

However, the broader cyclic Baxter-Bazhanov-Stroganov (BBS), or \(\tau^{(2)}\) model
is well established, possessing a Yang-Baxter structure and
separation-of-variables (SoV) formulations. For odd \(N\), the off-diagonal
Bethe ansatz (ODBA) also gives an inhomogeneous \(T\)--\(Q\)
description~\cite{BazhanovStroganov1990,Tarasov1992,vonGehlenEtAl2006,XuEtAl2015}.
For the periodic Baxter-Fendley chain, we make the Hamiltonian-level connection
with this transfer-matrix structure explicit.  The operator-valued matching
polynomial \(I(u)\) of the periodic current cycle is precisely the transfer
matrix \(t(u)\), and the Hamiltonian is its first coefficient, \(H=I_1\).
Once this identification is made, the known SoV and ODBA constructions can be
specialized to the clock chain.  The resulting SoV formulation is given in
Appendix~\ref{app:separation-of-variables-solution}, while for odd \(N\) on the stated
generic chart, the corresponding ODBA formulation is given in
Appendix~\ref{app:odba-finite-size-verification}.

Matching polynomials and related independence polynomials have recently
attracted increasing interest in the study of integrable quantum spin systems,
particularly in the characterization of free-parafermion solvability and the
construction of conserved charges~\cite{MannElmanWoodChapman2025,FukaiPozsgayVona2026}.
Here we extend the graph-theoretic independent-set approach to spectral analysis
beyond free-fermion and free-parafermion models, deriving explicit finite-size
spectral equations for the periodic Baxter-Fendley \(Z_N\) chain with \(N>2\).
The operator matching polynomial and its commuting coefficients are
defined for arbitrary couplings~\cite{MannElmanWoodChapman2025}. On the nonzero
odd-coupling chart, root-of-unity fusion maps every physical transfer
eigenvalue in a fixed charge sector to a common zero of $L-1$
polynomial equations in the nontrivial coefficients of the transfer
eigenvalue. Each equation has degree exactly $N$. We show that their
projective homogenizations have no common zero at infinity. 
B\'ezout's theorem then gives
$N^{L-1}=\dim{\cal H}_q$ finite solutions. We further show that the
normalized energy resultant of this polynomial system coincides
identically with the charge-sector characteristic polynomial. Thus
the finite spectral equations reproduce the complete charge-sector
energy spectrum with algebraic multiplicity. Although the BBS fusion
derivation is made initially on the nonzero odd-coupling chart, the
resultant identity extends polynomially to arbitrary physical
couplings.

A second main result identifies the ground-state branch.  For every
\(N,L\geq2\), positive reflection-symmetric bulk couplings, and every seam
coupling \(\mu\geq0\), we prove that \(H(\mu)\) has a real, algebraically
simple eigenvalue whose real part is strictly larger than that of every other
eigenvalue.  Its eigenspace lies in the zero charge sector.  The proof uses
reflection folding to represent the evolution by a strongly positive
completely positive semigroup on a finite-dimensional \(C^*\)-algebra, to which
the Perron-Frobenius theory for positive maps applies~\cite{EvansHoeghKrohn1978}.
Consequently, the known OBC ground state has a unique analytic continuation
along the entire nonnegative seam path, and this continuation remains the
ground state~\cite{Kato1995}.  For homogeneous couplings, the endpoint \(\mu=\lambda\) gives
the PBC ground state.

Figure~\ref{fig:prl-two-cycle-closure} summarizes this two-cycle
representation. The periodic seam turns the OBC matching path into the PBC
cycle \(C_{2L}\), while root-of-unity fusion produces the spectral-orbit cycle
\(C_N\) that closes the finite spectral problem.

As a first application, we consider the periodic homogeneous \(N=3\) chain.
The exact finite-size seam-response formula determines its exponential rate
for \(0<\lambda<1\), yielding a saddle-point criterion for the loss of
exponential suppression.  
Numerical solution of this criterion, together with duality, gives a reciprocal pair of candidate PBC critical couplings, whose physical significance is supported by the finite-size ground-state curvature. Their interpretation within the complete periodic phase diagram, including the nature of the intervening region, will be investigated in future work. This behavior is sharply distinct from OBC, where the positive-real quasienergy gap closes at $\lambda=1$.

The remainder of the paper is organized as follows.  Section
\ref{sec:preliminaries} establishes the matching-polynomial and cyclic-Weyl
preliminaries.  
Sections~\ref{sec:operator-matching-polynomial}--\ref{sec:root-of-unity-fusion-closure} construct the commuting transfer family
and its root-of-unity closure.  Section~\ref{sec:finite-spectral-system}
studies the resulting polynomial system, while Sections
\ref{sec:ground-state-branch} and \ref{sec:physical-branch-continuation}
select and continue the physical ground-state branch.  Sections
\ref{sec:seam-response-asymptotics} and
\ref{sec:homogeneous-n3-application} treat the seam response and the
homogeneous-chain application, with \(N=3\) as an example.  Section
\ref{sec:conclusion-discussion} concludes the paper.  Appendices
\ref{app:odba-finite-size-verification} and
\ref{app:separation-of-variables-solution} give the ODBA and SoV formulations,
respectively.

\section{Preliminaries}\label{sec:preliminaries}

\begin{definition}
\label{def:weighted-cycle}
Let \(C_n\), \(n\geq2\), be the cyclic graph with vertex set
\(\{1,\ldots,n\}\) and edges
\[
  e_j=\{j,j+1\},
  \qquad j=1,\ldots,n,
\]
where vertex indices are read modulo \(n\).  For \(n=2\), this means the
multigraph with two parallel indexed edges \(e_1\) and \(e_2\).
Assign a vertex weight \(v_j\) to vertex \(j\) and an edge weight \(w_j\)
to \(e_j\).  We denote the set of matchings of \(C_n\) by
\(\mathfrak M(C_n)\).  For \(F\in\mathfrak M(C_n)\), let \(V(F)\) be the
set of vertices incident to the edges in \(F\).  We choose vertex \(1\) as
a base point and read products in the order \(1,2,\ldots,n\).
\end{definition}

\begin{definition}
\label{def:weighted-matching-functional}
Let \(\mathcal R\) be an associative unital ring, let \(u\) and \(\xi\)
be central indeterminates, and let
\(v_j(u),w_j(u)\in\mathcal R[u]\).  The matrix
\begin{equation}
  \mathsf M[v_j,w_j](u;\xi):=
  \begin{pmatrix}
    v_j(u)&\xi w_j(u)\\
    1_{\mathcal R}&0
  \end{pmatrix}
  \label{eq:weighted-matching-matrix}
\end{equation}
generates the weighted matching polynomial of the cycle through the trace
of its ordered product, with the corresponding choice of edge
orientation~\cite{FacchiniLeroy2015}.  We denote this functional by
\begin{equation}
  Z_{C_n}^{(1)}\!\left(\mathsf M[v,w],u;\xi\right)
  :=
  \operatorname{Tr}_2\!\left[
    \prod_{j=1}^{n}\mathsf M[v_j,w_j](u;\xi)
  \right],
  \label{eq:based-weighted-matching-functional}
\end{equation}
where \(\operatorname{Tr}_2(A):=A_{00}+A_{11}\) is the diagonal sum of a
\(2\times2\) matrix.  The superscript records the base point, with no cyclicity
of \(\operatorname{Tr}_2\) assumed.
The first argument of \(Z_{C_n}^{(1)}\) denotes the family
\(\mathsf M[v_j,w_j]\) read in this based cyclic order.
\end{definition}

\begin{remark}
Equation~\eqref{eq:based-weighted-matching-functional} is a matrix-product representation with auxiliary
dimension two. In the operator-valued chain specialization below it
becomes a periodic matrix-product operator (MPO) of bond dimension
two. After pairing consecutive matching matrices as in 
Eq.~(\ref{eq:paired-matching-matrix}) and
performing the gauge transformation in Eq.~(\ref{eq:site-local-l-operator}), each local tensor
acts on a single clock site.
\end{remark}

The following definition is the periodic extension of Fendley's open-chain
current construction~\cite{Fendley2014}.

\begin{definition}
\label{def:periodic-current-family}
Fix \(N,L\geq2\), set \(\omega=e^{2\pi i/N}\), and let
\(\mathcal H=(\mathbb C^N)^{\otimes L}\).  For
\(\boldsymbol\lambda=(\lambda_1,\ldots,\lambda_{2L})\in\mathbb C^{2L}\),
define the periodic current family
\begin{equation}
\begin{aligned}
  h_{2j-1}=\lambda_{2j-1}\tau_j,\qquad 
  h_{2j}=\lambda_{2j}\sigma_j^\dagger\sigma_{j+1},\qquad 
  h_{n+2L}=h_n,
\end{aligned}
  \label{eq:cyclic-currents}
\end{equation}
where \(\tau_j\) is the unitary shift operator and \(\sigma_j\) is the
unitary clock operator on site \(j\), acting as
\[
  \tau_j\lvert s\rangle_j=\lvert s+1\rangle_j,
  \qquad
  \sigma_j\lvert s\rangle_j=\omega^s\lvert s\rangle_j,\qquad \sigma_{L+1}=\sigma_1,
  \qquad s\in\mathbb Z_N.
\]
They satisfy \(\tau_j^N=\sigma_j^N=\mathbf 1\),
\(\sigma_j\tau_j=\omega\tau_j\sigma_j\), and operators on distinct sites
commute.
\end{definition}

\begin{lemma}
\label{lem:cyclic-weyl-relations}
The current family in Definition~\ref{def:periodic-current-family} satisfies
\begin{equation}
  h_nh_{n+1}=\omega h_{n+1}h_n,
  \qquad
  h_n^N=\lambda_n^N\mathbf 1,
  \qquad
  [h_n,h_m]=0
  \quad\text{if }m\neq n\pm1\pmod{2L},
  \label{eq:cyclic-weyl-relations}
\end{equation}
for \(1\leq n,m\leq2L\).
\end{lemma}
This follows directly from the local Weyl relation and commutativity of operators on distinct sites.

\Needspace{12\baselineskip}
\begin{proposition}%
\label{prop:noncommutative-circular-trace}
For the data in Definitions~\ref{def:weighted-cycle}
and~\ref{def:weighted-matching-functional},
the functional \(Z_{C_n}^{(1)}\) is the sum over matchings
\(F\in\mathfrak M(C_n)\), with a factor \(\xi w_j(u)\) at position
\(j\) for each edge \(e_j\in F\) and a factor \(v_i(u)\) at position
\(i\) for each unmatched vertex.  Within each contribution, these factors
are multiplied in increasing vertex order \(1,2,\ldots,n\).
No commutativity among the weights is required.  If all weights commute,
the base-point label may be suppressed, and
\begin{equation}
\begin{gathered}
  Z_{C_n}\!\left(\mathsf M[v,w],u;\xi\right)
  =
  \operatorname{tr}\!\left[
    \prod_{j=1}^{n}\mathsf M[v_j,w_j](u;\xi)
  \right]
  \\
  =
  \sum_{F\in\mathfrak M(C_n)}\xi^{|F|}
  \prod_{e_j\in F}w_j(u)
  \prod_{i\notin V(F)}v_i(u).
\end{gathered}
  \label{eq:weighted-cycle-matching-trace}
\end{equation}
\end{proposition}

\begin{proof}
Write the row and column indices of the local matrices as \(0,1\).  Expanding
the diagonal sum of the fixed ordered product gives
\begin{equation}
  \operatorname{Tr}_2(\mathsf M_1\cdots\mathsf M_n)
  =\sum_{\substack{s_0,\ldots,s_{n-1}\in\{0,1\}\\ s_n=s_0}}
    \prod_{j=1}^{n}(\mathsf M_j)_{s_{j-1},s_j},
  \label{eq:closed-binary-path-expansion}
\end{equation}
where the factors on the right remain in increasing order of \(j\).  The only
nonzero local transitions are
\(0\to0\), \(0\to1\), and \(1\to0\), with respective weights
\(v_j(u)\), \(\xi w_j(u)\), and \(1_{\mathcal R}\).  Since the
\(1\to1\) entry vanishes, every \(0\to1\) transition at position \(j\) is
followed cyclically by a \(1\to0\) transition at position \(j+1\).
Consequently, the set
\begin{equation*}
  F=\{e_j:(s_{j-1},s_j)=(0,1)\}
\end{equation*}
is a matching of \(C_n\).  Conversely, every matching determines a unique
closed binary path of this form, including matchings that contain the seam
edge \(e_n\).  The corresponding ordered matrix-entry product contains
exactly the selected edge weights and unmatched vertex weights in the
stated order, with a factor \(\xi\) for each edge.  Summing over the
resulting bijection proves the matching expansion. No reordering of ring elements
and no cyclic invariance of the matrix trace is used.
\end{proof}

\section{The operator matching polynomial}
\label{sec:operator-matching-polynomial}

In this section, \(h_1,\ldots,h_{2L}\) denotes the periodic current
family of Definition~\ref{def:periodic-current-family}, which satisfies the
cyclic Weyl relations in Lemma~\ref{lem:cyclic-weyl-relations}.

\begin{remark}
\label{rem:oriented-frustration-cycle}
Arrange the currents \(h_1,\ldots,h_{2L}\) in this order around a circle, with
\(h_{2L}\) followed by \(h_1\).  Taking these currents as vertices, the cyclic
Weyl relations give the cyclically oriented frustration graph
\(\overrightarrow C_{2L}\).  For \(2L\geq4\), this graph is dipath-oriented
but is not an oriented indifference graph.  Thus the dipath-orientation
hypothesis for commuting independent set charges is satisfied, whereas the
oriented-indifference hypothesis that is sufficient for a free-parafermion
solution is not~\cite{MannElmanWoodChapman2025}.

For the matching construction below, take the canonically labelled cycle
\(C_{2L}\) with
\[
  e_j=\{j,j+1\},
  \qquad j\in\mathbb Z/(2L)\mathbb Z,
\]
and assign \(h_j\) to \(e_j\), in the same cyclic order, while every vertex
has weight \(\mathbf 1_{\mathcal H}\).  Its line graph is the current cycle
above under the identification \(e_j\leftrightarrow h_j\).  The corresponding
specialization of the matrix in Definition~\ref{def:weighted-matching-functional}
is
\begin{equation}
  \mathsf M_j(u)
  :=\mathsf M[\mathbf 1_{\mathcal H},h_j](u;-u)
  =
  \begin{pmatrix}
    \mathbf 1_{\mathcal H}&-u h_j\\
    \mathbf 1_{\mathcal H}&0
  \end{pmatrix}.
  \label{eq:current-matching-matrix}
\end{equation}
\end{remark}

\begin{definition}
\label{def:operator-matching-polynomial}
Let \(\mathfrak M_k(C_{2L})\) be the set of \(k\)-edge matchings.  For
\(F\in\mathfrak M_k(C_{2L})\), put
\begin{equation}
  h_F:=\prod_{e_j\in F}h_j,
  \qquad
  I_k:=\sum_{F\in\mathfrak M_k(C_{2L})}h_F,
  \qquad I_0:=\mathbf 1_{\mathcal H}.
  \label{eq:operator-matching-charges}
\end{equation}
The product defining $h_F$ is independent of its ordering, 
since the edges in a matching are pairwise nonincident 
and their current weights commute by Lemma~\ref{lem:cyclic-weyl-relations}.  
The operator matching polynomial is
\begin{equation}
\begin{aligned}
  I(u)
  &:={}
  Z_{C_{2L}}^{(1)}\!\left(
    \mathsf M[\mathbf 1_{\mathcal H},h_j],u;-u
  \right)
   =\sum_{k=0}^{L}(-u)^kI_k,\\
  H&:=I_1
   =\sum_{n=1}^{2L}h_n
   =\sum_{j=1}^{L}\left(
      \lambda_{2j-1}\tau_j
      +\lambda_{2j}\sigma_j^\dagger\sigma_{j+1}
    \right).
\end{aligned}
  \label{eq:operator-matching-polynomial}
\end{equation}
The Hamiltonian \(H\) in Eq.~\eqref{eq:operator-matching-polynomial} is the
periodic-boundary extension of the open Baxter-Fendley \(Z_N\) clock
chain~\cite{Baxter1989,Fendley2014}.
Since every matching monomial is order independent, this specialization is
independent of the chosen base point, and the superscript on \(Z_{C_{2L}}^{(1)}\)
may be suppressed.  It is the operator-valued analogue of the monomer-dimer
polynomial appearing in the Heilmann-Lieb theorem~\cite{HeilmannLieb1972},
with dimer activity \(-u h_j\) and unit monomer activity.
\end{definition}

Set
\begin{equation}
  \lambda_{\mathrm o}:=\prod_{j=1}^{L}\lambda_{2j-1},
  \qquad
  \lambda_{\mathrm e}:=\prod_{j=1}^{L}\lambda_{2j},
  \qquad
  \omega^{\mathcal P}:=\prod_{j=1}^{L}\tau_j,
  \label{eq:coupling-products-and-charge}
\end{equation}
where \(\omega^{\mathcal P}\) is the global \(Z_N\)-charge
operator~\cite{Fendley2014}.

The two perfect matchings of \(C_{2L}\) give the highest matching charge
\begin{equation}
  I_L=\lambda_{\mathrm o}\omega^{\mathcal P}
      +\lambda_{\mathrm e}\mathbf 1_{\mathcal H}.
  \label{eq:endpoint-matching-charges}
\end{equation}

\begin{theorem}
\label{thm:commuting-matching-charges}
For arbitrary couplings
\(\boldsymbol\lambda\in\mathbb C^{2L}\), the matching charges commute
pairwise:
\begin{equation}
  [I_k,I_\ell]=0,
  \qquad 0\leq k,\ell\leq L.
  \label{eq:commuting-matching-charges}
\end{equation}
Consequently,
\begin{equation}
  [I(\mu),I(\nu)]=0
  \quad\text{for all }\mu,\nu\in\mathbb C,
  \qquad
  [H,I_k]=0
  \quad\text{for }0\leq k\leq L.
  \label{eq:commuting-operator-matching-polynomials}
\end{equation}
\end{theorem}

\begin{proof}
First suppose that all couplings are real and nonzero.  Under the
identification \(e_j\leftrightarrow h_j\), the matchings of \(C_{2L}\) are the
independent sets of its line graph.  This line graph is again a cycle, and the
cyclic Weyl relations give it precisely the dipath orientation described in
Remark~\ref{rem:oriented-frustration-cycle}.  Hence \(I_k\) is the \(k\)-th
independent set charge in the sense of Mann, Elman, Wood, and Chapman.  For a
qudit Hamiltonian with a dipath-oriented frustration graph, the independent
set charges commute pairwise~\cite{MannElmanWoodChapman2025}.  Therefore
Eq.~\eqref{eq:commuting-matching-charges} follows.

For general complex couplings, every matrix entry of \([I_k,I_\ell]\) is a
polynomial in \(\lambda_1,\ldots,\lambda_{2L}\).  These polynomials vanish 
for all real nonzero couplings. Successively regarding each entry as a 
one-variable polynomial shows that it vanishes identically.
Thus Eq.~\eqref{eq:commuting-matching-charges} holds for all complex couplings,
including vanishing ones.  Finally,
\begin{equation*}
  [I(\mu),I(\nu)]
  =\sum_{k,\ell=0}^{L}
    (-\mu)^k(-\nu)^\ell[I_k,I_\ell]=0,
\end{equation*}
and \(H=I_1\) gives the remaining assertion.
\end{proof}

Related direct constructions of commuting charges for this exchange-algebra class, 
including periodic cases, are given in Ref.~\cite{Alcaraz2026Nonhomogeneous}.

Matchings of \(C_{2L}\) are independent sets of its line graph, which is
again \(C_{2L}\).  Accordingly, \(I(u)\) is also an operator-valued
specialization of the multivariate independence polynomial
\cite{RadchenkoRodriguezVillegas2021}.  Independence-polynomial transfer
operators have previously been used to construct commuting charges for Pauli
Hamiltonians with claw-free frustration graphs and explicit free-fermion
spectra in the even-hole-free subclass~\cite{ElmanChapmanFlammia2021}.

\section{Yang-Baxter structure}
\label{sec:yang-baxter-structure}

For \(N>2\), the cyclic Weyl algebra is non-fermionic, and the free-fermion
reduction discussed above is not available.  Instead, the matching matrices
admit a direct Yang-Baxter construction.  In the closed-path expansion of
Proposition~\ref{prop:noncommutative-circular-trace}, the zero lower-right
entry of each matrix excludes incident occupied edges, an occupied edge
\(e_n\) contributes \(-u h_n\), and the auxiliary diagonal sum closes the
path across the periodic seam.  Pairing consecutive matrices exposes a local
structure.  Its root-of-unity consequences will be treated by fusion rather
than by a free-fermion reduction.

\begin{definition}
\label{def:site-local-l-operator}
Let \(\mathcal V_a\simeq\mathbb C^2\) be an auxiliary space, and write
\(\mathsf M_a[v,w](u;\xi)\) for the matrix in
Definition~\ref{def:weighted-matching-functional} acting in
\(\mathcal V_a\).  For each clock site \(j\), pair the matrices carrying
\(h_{2j-1}\) and \(h_{2j}\):
\begin{equation}
\begin{aligned}
  P_{a,j}(u)
  &:=
  \mathsf M_a[\mathbf 1_{\mathcal H},h_{2j-1}](u;-u)
  \mathsf M_a[\mathbf 1_{\mathcal H},h_{2j}](u;-u)\\
  &=
  \begin{pmatrix}
    \mathbf 1_{\mathcal H}-u\lambda_{2j-1}\tau_j&
    -u\lambda_{2j}\sigma_j^\dagger\sigma_{j+1}\\
    \mathbf 1_{\mathcal H}&
    -u\lambda_{2j}\sigma_j^\dagger\sigma_{j+1}
  \end{pmatrix}_{\!a}.
\end{aligned}
  \label{eq:paired-matching-matrix}
\end{equation}
The entries of \(P_{a,j}(u)\) act on clock sites \(j\) and \(j+1\). For 
\(j=L\), these are the two sites adjacent to the periodic seam.  Define the
periodic auxiliary-space gauge
\begin{equation}
  g_{a,j}:=
  \begin{pmatrix}
    \mathbf 1_{\mathcal H}&0\\
    0&\sigma_j
  \end{pmatrix}_{\!a},
  \qquad
  g_{a,j}^{-1}=
  \begin{pmatrix}
    \mathbf 1_{\mathcal H}&0\\
    0&\sigma_j^\dagger
  \end{pmatrix}_{\!a},
  \qquad
  g_{a,L+1}=g_{a,1},
  \label{eq:periodic-auxiliary-gauge}
\end{equation}
and set
\begin{equation}
\begin{aligned}
  L_{a,j}(u)
  &:=
  g_{a,j}P_{a,j}(u)g_{a,j+1}^{-1}\\
  &=
  \begin{pmatrix}
    \mathbf 1_{\mathcal H}-u\lambda_{2j-1}\tau_j&
    -u\lambda_{2j}\sigma_j^\dagger\\
    \sigma_j&
    -u\lambda_{2j}\mathbf 1_{\mathcal H}
  \end{pmatrix}_{\!a}.
\end{aligned}
  \label{eq:site-local-l-operator}
\end{equation}
Thus every quantum-space entry of \(L_{a,j}(u)\) acts only on clock site
\(j\).
\end{definition}

Let \(\mathcal V_b\simeq\mathbb C^2\) be a second auxiliary space.  In the
ordered basis
\((e_0\otimes e_0,e_0\otimes e_1,e_1\otimes e_0,e_1\otimes e_1)\)
of \(\mathcal V_a\otimes\mathcal V_b\), define
\begin{equation}
  R_{ab}(\mu,\nu):=
  \begin{pmatrix}
    \mu-\omega\nu&0&0&0\\
    0&\omega(\mu-\nu)&\mu(1-\omega)&0\\
    0&\nu(1-\omega)&\mu-\nu&0\\
    0&0&0&\mu-\omega\nu
  \end{pmatrix}_{\!ab}.
  \label{eq:twisted-six-vertex-r-matrix}
\end{equation}

\begin{proposition}
\label{prop:local-rll-relation}
For every \(j=1,\ldots,L\) and every \(\mu,\nu\in\mathbb C\), the
site-local operator in Eq.~\eqref{eq:site-local-l-operator} satisfies
\begin{equation}
  R_{ab}(\mu,\nu)L_{a,j}(\mu)L_{b,j}(\nu)
  =
  L_{b,j}(\nu)L_{a,j}(\mu)R_{ab}(\mu,\nu).
  \label{eq:local-rll-relation}
\end{equation}
\end{proposition}

\begin{proof} 
For \(\lambda_{2j-1}\neq0\), the operator \(L_{a,j}(u)\) in
Eq.~\eqref{eq:site-local-l-operator} is a special case of the cyclic BBS
\(L\)-operator (the explicit parameter specialization is given in 
Eq.~\eqref{eq:app-a-bbs-specialization}).  
The cyclic BBS \(L\)-operator satisfies the local RLL
relation with the root-of-unity twisted six-vertex
\(R\)-matrix~\cite{vonGehlenEtAl2006}.  Since both sides of
Eq.~\eqref{eq:local-rll-relation} depend polynomially on the couplings, the
identity extends to \(\lambda_{2j-1}=0\).
\end{proof}

\begin{definition}
\label{def:monodromy-transfer-matrix}
The monodromy matrix and transfer matrix associated with the local operators
in Eq.~\eqref{eq:site-local-l-operator} are
\begin{equation}
\begin{aligned}
  T_a(u)
  &:=L_{a,1}(u)L_{a,2}(u)\cdots L_{a,L}(u)
    =
    \begin{pmatrix}
      A(u)&B(u)\\
      C(u)&D(u)
    \end{pmatrix}_{\!a},\\
  t(u)&:=\operatorname{Tr}_a T_a(u)=A(u)+D(u),
\end{aligned}
  \label{eq:monodromy-transfer-matrix}
\end{equation}
where \(\operatorname{Tr}_a\) denotes the diagonal sum over the auxiliary
space \(\mathcal V_a\).
\end{definition}

\begin{lemma}
\label{lem:transfer-matching-identity}
The transfer matrix is the operator matching polynomial
\begin{equation}
  t(u)=I(u).
  \label{eq:transfer-matching-identity}
\end{equation}
\end{lemma}

\begin{proof}
Let
\begin{equation}
\begin{aligned}
  \mathcal U_a(u)
  &:={}
  P_{a,1}(u)P_{a,2}(u)\cdots P_{a,L}(u)\\
  &=\mathsf M_1(u)\mathsf M_2(u)\cdots\mathsf M_{2L}(u)
   =
   \begin{pmatrix}
     \mathcal U_{00}(u)&\mathcal U_{01}(u)\\
     \mathcal U_{10}(u)&\mathcal U_{11}(u)
   \end{pmatrix}_{\!a}.
\end{aligned}
  \label{eq:ungauged-monodromy}
\end{equation}
Here \(\mathsf M_n(u)\) is the current matching matrix defined in
Eq.~\eqref{eq:current-matching-matrix}.
The adjacent gauges cancel in their displayed order, without being commuted
through any quantum-space operator.  Since \(g_{a,L+1}=g_{a,1}\),
\begin{equation}
  T_a(u)=g_{a,1}\mathcal U_a(u)g_{a,1}^{-1}.
  \label{eq:gauge-telescoping}
\end{equation}

In the closed-path expansion of \(\mathcal U_{11}(u)\), the path begins in
auxiliary state \(1\).  The second row of \(\mathsf M_1(u)\) contains no
entry proportional to \(h_1\), and therefore no monomial in
\(\mathcal U_{11}(u)\) contains \(h_1\).  Every other current commutes with
\(\sigma_1\). In particular,
\(h_2\propto\sigma_1^\dagger\sigma_2\) and
\(h_{2L}\propto\sigma_L^\dagger\sigma_1\), while all remaining currents
either act away from site \(1\) or contain \(\tau_j\) with \(j\neq1\).  Hence
\begin{equation}
  [\mathcal U_{11}(u),\sigma_1]=0.
  \label{eq:lower-right-boundary-commutator}
\end{equation}
It follows directly from Eq.~\eqref{eq:gauge-telescoping} that
\begin{equation}
\begin{aligned}
  \operatorname{Tr}_aT_a(u)
  &=\mathcal U_{00}(u)
    +\sigma_1\mathcal U_{11}(u)\sigma_1^\dagger\\
  &=\mathcal U_{00}(u)+\mathcal U_{11}(u)
   =\operatorname{Tr}_a\mathcal U_a(u).
\end{aligned}
  \label{eq:residual-gauge-trace}
\end{equation}
Proposition~\ref{prop:noncommutative-circular-trace} and
Eq.~\eqref{eq:operator-matching-polynomial} identify the last expression
with \(I(u)\).  This proves Eq.~\eqref{eq:transfer-matching-identity} without
using cyclicity of an operator-valued auxiliary trace.
\end{proof}

\begin{theorem}
\label{thm:global-rtt-relation}
For all \(\mu,\nu\in\mathbb C\), the monodromy matrix satisfies the RTT
relation
\begin{equation}
  R_{ab}(\mu,\nu)T_a(\mu)T_b(\nu)
  =T_b(\nu)T_a(\mu)R_{ab}(\mu,\nu).
  \label{eq:global-rtt-relation}
\end{equation}
Consequently, the transfer matrices commute:
\begin{equation}
  [t(\mu),t(\nu)]=0.
  \label{eq:transfer-matrix-commutativity}
\end{equation}
\end{theorem}

\begin{proof}
Operators belonging to different quantum sites commute, so multiplying
the local RLL relations in Eq.~(\ref{eq:local-rll-relation}) in site order gives the RTT relation
(\ref{eq:global-rtt-relation}). For generic $\mu,\nu$ the numerical matrix $R_{ab}(\mu,\nu)$ is
invertible. Multiplying Eq.~(\ref{eq:global-rtt-relation}) by $R_{ab}^{-1}$ and taking the
diagonal sum over both auxiliary spaces gives
$t(\mu)t(\nu)=t(\nu)t(\mu)$. Since both sides are polynomial in
$\mu$ and $\nu$, the identity extends to all spectral parameters.
\end{proof}

It follows that \(I(u)=t(u)\) is the corresponding
transfer matrix in the present clock-chain specialization.  Together with
\(H=I_1\), Theorem~\ref{thm:global-rtt-relation} gives an independent
Yang-Baxter derivation of the commuting family in
Theorem~\ref{thm:commuting-matching-charges}. 
Together with $H=I_1$, Theorem~\ref{thm:global-rtt-relation} gives a Yang-Baxter realization of the commuting family in Theorem~\ref{thm:commuting-matching-charges} and makes the transfer-matrix structure of the periodic clock chain explicit.

For \(u\neq0\), the matrix \(R(\omega u,u)\) has rank one.  For local
operators satisfying the RLL relation, the corresponding quantum determinant
is multiplicative under ordered ultralocal products, as established by the
rank-one fusion construction~\cite{IzerginKorepin1981,vonGehlenEtAl2006}.

\begin{definition}
\label{def:quantum-determinant}
Within the Yang-Baxter algebra determined by
Eq.~\eqref{eq:twisted-six-vertex-r-matrix}, let \(X_a(u)\) be a local
\(L\)-operator or a monodromy matrix satisfying the corresponding RLL or RTT
relation.  Writing
\(
X_a(u)=\bigl(\begin{smallmatrix}
X_{00}(u)&X_{01}(u)\\ X_{10}(u)&X_{11}(u)
\end{smallmatrix}\bigr)_a
\), define
\begin{equation}
  \det\nolimits_qX_a(u)
  :=X_{00}(\omega u)X_{11}(u)-X_{10}(\omega u)X_{01}(u).
  \label{eq:quantum-determinant-definition}
\end{equation}
In particular, for \(r\in\mathbb Z_N\),
\begin{equation}
  \det\nolimits_qT_a(\omega^{r-1}u)
  =A(\omega^ru)D(\omega^{r-1}u)
   -C(\omega^ru)B(\omega^{r-1}u).
  \label{eq:shifted-monodromy-quantum-determinant}
\end{equation}
Set
\begin{equation}
  \lambda_{\mathrm{tot}}
  :=\prod_{n=1}^{2L}\lambda_n
  =\lambda_{\mathrm o}\lambda_{\mathrm e}.
  \label{eq:total-coupling-product}
\end{equation}
\end{definition}

\begin{proposition}
\label{prop:quantum-determinant}
For every \(j=1,\ldots,L\) and \(r\in\mathbb Z_N\),
\begin{equation}
  \det\nolimits_qL_{a,j}(\omega^{r-1}u)
  =\omega^{2r-1}u^2
    \lambda_{2j-1}\lambda_{2j}\tau_j.
  \label{eq:local-quantum-determinant}
\end{equation}
The quantum determinant of the monodromy is
\begin{equation}
  \det\nolimits_qT_a(\omega^{r-1}u)
  =\omega^{L(2r-1)}u^{2L}
    \lambda_{\mathrm{tot}}\,\omega^{\mathcal P}.
  \label{eq:monodromy-quantum-determinant}
\end{equation}
\end{proposition}

\begin{proof}
From Eq.~(\ref{eq:site-local-l-operator}),
\[
A_j(u)=1-u\lambda_{2j-1}\tau_j,\quad
B_j(u)=-u\lambda_{2j}\sigma_j^\dagger,\quad
C_j(u)=\sigma_j,\quad
D_j(u)=-u\lambda_{2j}.
\]
Hence
\begin{align*}
\det{}_q L_{a,j}(\omega^{r-1}u)
&=
A_j(\omega^r u)D_j(\omega^{r-1}u)
-
C_j(\omega^r u)B_j(\omega^{r-1}u)\\
&=
\omega^{2r-1}u^2
\lambda_{2j-1}\lambda_{2j}\tau_j ,
\end{align*}
because the two terms linear in $u$ cancel.
Multiplicativity of the quantum determinant for the ordered
ultralocal product then gives
\[
\det{}_q T_a(\omega^{r-1}u)
=
\prod_{j=1}^L
\det{}_q L_{a,j}(\omega^{r-1}u)
=
\omega^{L(2r-1)}u^{2L}\lambda_{\rm tot}\omega^P .
\]
\end{proof}

\begin{definition}
\label{def:charge-sectors}
The operator \(\omega^{\mathcal P}\) in
Eq.~\eqref{eq:coupling-products-and-charge} is the global \(Z_N\)-charge of
the clock chain~\cite{Fendley2014}.  For \(q\in\mathbb Z_N\), define
\begin{equation}
  \mathcal H_q
  :=\bigl\{\lvert\Psi\rangle\in\mathcal H:
       \omega^{\mathcal P}\lvert\Psi\rangle
       =\omega^q\lvert\Psi\rangle\bigr\}.
  \label{eq:charge-sector-definition}
\end{equation}
\end{definition}

\begin{lemma}
\label{lem:charge-sector-quantum-determinant}
The charge commutes with the full transfer family, and the Hilbert space has
the invariant decomposition
\begin{equation}
  [t(u),\omega^{\mathcal P}]=0,
  \qquad
  \mathcal H=\bigoplus_{q\in\mathbb Z_N}\mathcal H_q,
  \qquad
  \dim\mathcal H_q=N^{L-1}.
  \label{eq:charge-sector-decomposition}
\end{equation}
On \(\mathcal H_q\), Eq.~\eqref{eq:monodromy-quantum-determinant} becomes
\begin{equation}
\begin{aligned}
  \alpha_{q,r}(u)\mathbf 1_{\mathcal H_q}
  &:={}
  \left.\det\nolimits_qT_a(\omega^{r-1}u)\right|_{\mathcal H_q},\\
  \alpha_{q,r}(u)
  &={}
  \omega^{q+L(2r-1)}\lambda_{\mathrm{tot}}u^{2L}.
\end{aligned}
  \label{eq:sector-quantum-determinant}
\end{equation}
\end{lemma}

\begin{proof}
The charge commutes with every odd current \(h_{2j-1}\).  For an even current
\(h_{2j}=\lambda_{2j}\sigma_j^\dagger\sigma_{j+1}\), the two Weyl phases
arising from \(\tau_j\) and \(\tau_{j+1}\) cancel.  Hence
\([\omega^{\mathcal P},h_n]=0\) for every \(n\), and
Eq.~\eqref{eq:transfer-matching-identity} gives
\([t(u),\omega^{\mathcal P}]=0\).

Since \((\omega^{\mathcal P})^N=\mathbf 1_{\mathcal H}\), its eigenspaces
give the stated direct sum.  In a simultaneous eigenbasis of
\(\tau_1,\ldots,\tau_L\), fixing the product eigenvalue fixes one of the
\(L\) labels after the other \(L-1\) have been chosen.  Thus every sector has
dimension \(N^{L-1}\).  Restricting
Eq.~\eqref{eq:monodromy-quantum-determinant} to \(\mathcal H_q\) proves
Eq.~\eqref{eq:sector-quantum-determinant}.
\end{proof}

In \(\det\nolimits_q\), the subscript \(q\) stands for \emph{quantum} and is
unrelated to the charge-sector label \(q\).

\section{Root-of-unity fusion and closure}
\label{sec:root-of-unity-fusion-closure}

\begin{definition}
\label{def:spectral-fusion-polynomial}
Fix \(q\in\mathbb Z_N\), and let
\(\lvert\Psi_q\rangle\in\mathcal H_q\) be a nonzero common eigenvector of
the commuting transfer family.  Suppressing any additional joint-spectral
label, define \(\Lambda_q(u)\) by
\begin{equation}
  t(u)\lvert\Psi_q\rangle
  =\Lambda_q(u)\lvert\Psi_q\rangle.
  \label{eq:transfer-eigenvalue}
\end{equation}
Since \(t(0)=I(0)=\mathbf 1_{\mathcal H}\), one has
\(\Lambda_q(0)=1\).  With the indices of
\(\alpha_{q,r}(u)\) read modulo \(N\), define
\begin{equation}
  \Phi_q[\Lambda_q](u)
  :=Z_{C_N}\!\left(
    \mathsf M[\Lambda_q(\omega^{N-r}u),\alpha_{q,N-r}(u)],
    u;-1
  \right),
  \qquad r=1,\ldots,N,
  \label{eq:spectral-matching-polynomial}
\end{equation}
where the first argument denotes the displayed family of \(N\) matrices in
increasing order of \(r\).
\end{definition}

The fusion construction for the cyclic six-vertex Yang-Baxter algebra is
controlled by degeneration points of the six-vertex \(R\)-matrix and is
realized by products of spectrally shifted \(L\)-operators
\cite{BazhanovStroganov1990}.  The resulting fused transfer hierarchy obeys
a second-order recurrence in the shifted spectral parameter
\cite{vonGehlenEtAl2006}.  At \(\omega^N=1\), the shifts form the finite orbit
\(u,\omega u,\ldots,\omega^{N-1}u\). Proposition~\ref{prop:noncommutative-circular-trace}
then identifies Eq.~\eqref{eq:spectral-matching-polynomial} as the
matching-polynomial form of the corresponding root-of-unity fusion
polynomial.

\begin{definition}
\label{def:root-of-unity-average}
Let \(O(u)\) be an operator-valued polynomial such that
\([O(\omega^m u),O(\omega^{m'}u)]=0\) for all
\(m,m'\in\mathbb Z_N\).  Its root-of-unity average is
\begin{equation}
  \mathcal O(u^N)
  :=\langle O\rangle(u^N)
  :=\prod_{m=0}^{N-1}O(\omega^m u).
  \label{eq:root-of-unity-average}
\end{equation}
The construction is the cyclic averaging introduced for root-of-unity
monodromy matrices by Tarasov~\cite{Tarasov1992}.  The commutativity
assumption makes the product independent of its ordering.  It is invariant
under \(u\mapsto\omega u\), and hence is a polynomial in \(u^N\).
\end{definition}

\begin{lemma}
\label{lem:classical-local-l-operator}
The four entry families of the local operator
\(L_{a,j}(u)\) in Eq.~\eqref{eq:site-local-l-operator} satisfy the
commutativity condition in Definition~\ref{def:root-of-unity-average}.
Their entrywise averages form the scalar matrix
\begin{equation}
\begin{aligned}
  \bigl[\mathcal L_{a,j}(u^N)\bigr]_{\alpha\beta}
  &:=\left\langle
       \bigl[L_{a,j}\bigr]_{\alpha\beta}
     \right\rangle(u^N),\\
  \mathcal L_{a,j}(u^N)
  &=
  \begin{pmatrix}
    1-\lambda_{2j-1}^{N}u^N&-\lambda_{2j}^{N}u^N\\
    1&-\lambda_{2j}^{N}u^N
  \end{pmatrix}_{\!a}.
\end{aligned}
  \label{eq:classical-clock-lax-operator}
\end{equation}
Moreover,
\begin{equation}
  \mathcal L_{a,j}(u^N)
  =\mathsf M[1,\lambda_{2j-1}^N](u;-u^N)
   \mathsf M[1,\lambda_{2j}^N](u;-u^N).
  \label{eq:classical-lax-factorization}
\end{equation}
\end{lemma}

\begin{proof}
Each entry of \(L_{a,j}(u)\) is a polynomial in one fixed Weyl operator, so
its spectrally shifted values commute.  The identities
\begin{equation*}
  \prod_{m=0}^{N-1}(1-\omega^m x)=1-x^N,
  \qquad
  \prod_{m=0}^{N-1}\omega^m=(-1)^{N-1},
\end{equation*}
together with \(\sigma_j^N=\tau_j^N=\mathbf 1\), give the four entries in
Eq.~\eqref{eq:classical-clock-lax-operator}.  In particular, the two entries
proportional to \(-u\) acquire the sign
\((-1)^N(-1)^{N-1}=-1\).  Hence all Weyl operators disappear from the
average.  Multiplication of the two scalar matching matrices gives
Eq.~\eqref{eq:classical-lax-factorization}.
\end{proof}

\begin{definition}
\label{def:classical-current-matching-polynomial}
Reading the weights \(\lambda_n^N\), \(n=1,\ldots,2L\), in cyclic order,
define
\begin{equation}
  \mathcal Z_{\boldsymbol\lambda}^{(N)}(u^N)
  :=Z_{C_{2L}}\!\left(\mathsf M[1,\lambda_n^N],u;-u^N\right)
  =\operatorname{tr}_a\!\left[\prod_{j=1}^{L}\mathcal L_{a,j}(u^N)\right].
  \label{eq:classical-current-matching}
\end{equation}
The second equality follows from
Eq.~\eqref{eq:classical-lax-factorization} and the scalar case of
Proposition~\ref{prop:noncommutative-circular-trace}.
\end{definition}

\begin{lemma}
\label{lem:classical-current-subset-expansion}
For \(B\subseteq\{1,\ldots,L\}\), put
\(B+1:=\{j+1\pmod L:j\in B\}\).  Then
\begin{equation}
  \mathcal Z_{\boldsymbol\lambda}^{(N)}(u^N)
  =\sum_{B\subseteq\{1,\ldots,L\}}(-u^N)^{|B|}
       \prod_{j\in B}\lambda_{2j}^N
       \prod_{k\notin B\cup(B+1)}
       \bigl(1-u^N\lambda_{2k-1}^N\bigr).
  \label{eq:classical-current-subset}
\end{equation}
\end{lemma}

\begin{proof}
Fix the set \(B\) of even edges \(e_{2j}\) belonging to a matching of
\(C_{2L}\).  These edges contribute
\((-u^N)^{|B|}\prod_{j\in B}\lambda_{2j}^N\) and exclude the adjacent odd
edges indexed by \(B\cup(B+1)\).  Every remaining odd edge may be chosen
independently, and summing over those choices yields the last product in
Eq.~\eqref{eq:classical-current-subset}.  Summing over \(B\) proves the
formula.
\end{proof}

\begin{remark}
\label{rem:alternating-unit-edge-specialization}
If \(\lambda_{2j-1}=1\) for every \(j\),
Eq.~\eqref{eq:classical-current-subset} reduces to the alternating
unit-edge form.
\end{remark}

For the local operator in Eq.~\eqref{eq:site-local-l-operator}, the
determinant polynomial entering
the fusion hierarchy is
\(z(u)=\omega^{-L}\lambda_{\mathrm{tot}}u^{2L}\).  Consequently,
\begin{equation*}
  \omega^qz(\omega^ru)=\alpha_{q,r}(u),
  \qquad r\in\mathbb Z_N.
\end{equation*}
For the fixed eigenvalue \(\Lambda_q(u)\), let the scalar fused-transfer
polynomials be normalized by
\begin{equation}
\begin{gathered}
  \tau_q^{(0)}(u)=0,
  \qquad
  \tau_q^{(1)}(u)=1,\\
  \tau_q^{(j+1)}(u)
  =\Lambda_q(\omega^{j-1}u)\tau_q^{(j)}(u)
   -\alpha_{q,j-1}(u)\tau_q^{(j-1)}(u),
  \qquad j=1,\ldots,N.
\end{gathered}
  \label{eq:fused-transfer-recurrence}
\end{equation}
In particular, \(\tau_q^{(2)}(u)=\Lambda_q(u)\).  On the chart
\(\lambda_{2j-1}\neq0\), \(j=1,\ldots,L\), the root-of-unity truncation
identity is
\begin{equation}
  \tau_q^{(N+1)}(u)
  -\alpha_{q,0}(u)\tau_q^{(N-1)}(\omega u)
  =\operatorname{tr}_a\!\left[
      \prod_{j=1}^{L}\mathcal L_{a,j}(u^N)
    \right]
  =\mathcal Z_{\boldsymbol\lambda}^{(N)}(u^N).
  \label{eq:root-of-unity-truncation}
\end{equation}
The scalar fused-transfer polynomials in
Eq.~\eqref{eq:fused-transfer-recurrence} satisfy the root-of-unity truncation
identity in Eq.~\eqref{eq:root-of-unity-truncation}
\cite{BazhanovStroganov1990,Tarasov1992,Baxter2004,vonGehlenEtAl2006}.
For odd \(N\), the corresponding shifted higher-spin formulation satisfies
the same truncation relation~\cite{XuEtAl2015}.  The second equality follows
from Definition~\ref{def:classical-current-matching-polynomial}.
The fusion and truncation identities are polynomial in the BBS parameters, 
so the generic identities extend to the specialization in 
Eq.~(\ref{eq:app-a-bbs-specialization}).

\begin{theorem}
\label{thm:matching-polynomial-closure}
Assume that \(\lambda_{2j-1}\neq0\) for \(j=1,\ldots,L\).  For every
\(q\in\mathbb Z_N\) and every transfer eigenvalue \(\Lambda_q(u)\) arising
from a common transfer-family eigenvector in \(\mathcal H_q\),
\begin{equation}
  \Phi_q[\Lambda_q](u)
  =\mathcal Z_{\boldsymbol\lambda}^{(N)}(u^N)
  \qquad\text{in }\mathbb C[u].
  \label{eq:matching-polynomial-closure}
\end{equation}
\end{theorem}

\begin{proof}
The polynomial \(\tau_q^{(N+1)}(u)\) satisfies the root-of-unity truncation
identity in Eq.~\eqref{eq:root-of-unity-truncation}
\cite{Baxter2004,vonGehlenEtAl2006}.
The relation \(\omega^qz(\omega^ru)=\alpha_{q,r}(u)\) identifies the
quantum-determinant coefficients in Eq.~\eqref{eq:fused-transfer-recurrence}
with the edge weights of
Definition~\ref{def:spectral-fusion-polynomial}.  A direct induction on
Eq.~\eqref{eq:fused-transfer-recurrence}, written in companion-matrix form and
with the spectral shifts ordered as \(r=1,\ldots,N\), gives
\begin{equation*}
  \Phi_q[\Lambda_q](u)
  =\tau_q^{(N+1)}(u)
   -\alpha_{q,0}(u)\tau_q^{(N-1)}(\omega u).
\end{equation*}
Equation~\eqref{eq:root-of-unity-truncation} now gives
Eq.~\eqref{eq:matching-polynomial-closure}.
\end{proof}

\Needspace{0.30\textheight}
\begin{figure}[h!]
  \centering
  \resizebox{\textwidth}{!}{%
  \begin{tikzpicture}[
    graph vertex/.style={circle,draw,fill=white,minimum size=2.1mm,
      inner sep=0pt,line width=0.55pt},
    panel/.style={font=\footnotesize\mdseries,align=center},
    small label/.style={font=\footnotesize,align=center},
    graph edge/.style={line width=0.75pt,shorten <=2.5pt,shorten >=2.5pt},
    directed edge/.style={graph edge,
      -{Latex[length=1.7mm,width=1.1mm]}},
    flow/.style={line width=0.8pt,-{Latex[length=2.0mm,width=1.25mm]}},
    closure/.style={<->,>=Latex,line width=0.8pt}
  ]
    \node[panel] at (1.85,-1.70) {(a) OBC $Z_N$:\\matching path $P_{2L}$};
    \node[graph vertex] (p0) at (0.00,0.35) {};
    \node[graph vertex] (p1) at (0.65,0.35) {};
    \node[graph vertex] (p2) at (1.30,0.35) {};
    \node at (1.85,0.35) {$\cdots$};
    \node[graph vertex] (p3) at (2.40,0.35) {};
    \node[graph vertex] (p4) at (3.05,0.35) {};
    \node[graph vertex] (p5) at (3.70,0.35) {};
    \draw[graph edge] (p0)--(p1) node[midway,above=4pt,small label] {$h_1$};
    \draw[graph edge] (p1)--(p2) node[midway,below=4pt,small label] {$h_2$};
    \draw[line width=0.75pt] (p2)--(1.58,0.35);
    \draw[line width=0.75pt] (2.12,0.35)--(p3);
    \draw[graph edge] (p3)--(p4)
      node[midway,above=4pt,small label] {$h_{2L-2}$};
    \draw[graph edge] (p4)--(p5)
      node[midway,below=4pt,small label] {$h_{2L-1}$};
    \node[small label] at (1.85,-0.78)
      {line graph: current path $P_{2L-1}$\\independent quasienergies};

    \draw[flow] (4.12,0.35)--(5.50,0.35)
      node[midway,above=4pt,small label]{add seam $h_{2L}$};

    \begin{scope}[shift={(7.00,0.35)}]
      \node[panel] at (0,-2.05) {(b) PBC $Z_N$:\\matching cycle $C_{2L}$};
      \foreach \name/\angle in {c1/90,c2/30,c3/-30,c4/-90,c5/-150,c6/150}
        \node[graph vertex] (\name) at (\angle:0.98) {};
      \draw[directed edge] (c1)--(c2);
      \draw[directed edge] (c2)--(c3);
      \draw[line width=0.75pt,shorten <=2.5pt]
        (c3)--($(c3)!0.30!(c4)$);
      \foreach \t in {0.42,0.50,0.58}
        \fill ($(c3)!\t!(c4)$) circle[radius=0.45pt];
      \draw[line width=0.75pt,shorten >=2.5pt]
        ($(c3)!0.70!(c4)$)--(c4);
      \draw[directed edge] (c4)--(c5);
      \draw[directed edge] (c5)--(c6);
      \draw[directed edge] (c6)--(c1);
      \node[small label] at (60:1.26) {$h_1$};
      \node[small label] at (1.25,-0.34) {$h_2$};
      \node[small label] at (-1.38,-0.34) {$h_{2L-1}$};
      \node[small label] at (120:1.32) {$h_{2L}$};
      \node[small label] at (0,-1.48)
        {$I(u)$};
    \end{scope}

    \draw[closure] (8.65,0.35)--(10.95,0.35)
      node[midway,above=6pt,small label]{matching-polynomial\\closure};
    \node[small label] at (9.80,-0.78)
      {$\Phi_q[\Lambda_q](u)=\mathcal Z_{\boldsymbol\lambda}^{(N)}(u^N)$};

    \begin{scope}[shift={(12.45,0.35)}]
      \node[panel] at (0,-2.05) {(c) spectral orbit:\\matching cycle $C_N$};
      \foreach \name/\angle in {s1/90,s2/18,s3/-54,s4/-126,s5/162}
        \node[graph vertex] (\name) at (\angle:0.98) {};
      \draw[directed edge] (s1)--(s2);
      \draw[directed edge] (s2)--(s3);
      \draw[line width=0.75pt,shorten <=2.5pt]
        (s3)--($(s3)!0.30!(s4)$);
      \foreach \t in {0.42,0.50,0.58}
        \fill ($(s3)!\t!(s4)$) circle[radius=0.45pt];
      \draw[line width=0.75pt,shorten >=2.5pt]
        ($(s3)!0.70!(s4)$)--(s4);
      \draw[directed edge] (s4)--(s5);
      \draw[directed edge] (s5)--(s1);
      \node[small label] at (0,0.02) {$\Lambda_q(\omega^r u)$};
      \node[small label] at (0,-1.48) {$\Phi_q[\Lambda_q](u)$};
    \end{scope}
  \end{tikzpicture}%
  }
  \caption{Periodic closure as a two-cycle matching problem.  (a) The OBC
  auxiliary matching path $P_{2L}$ (line graph: current/frustration path
  $P_{2L-1}$) underlies the independent quasienergies.  Adding the seam
  $h_{2L}$ gives the PBC $Z_N$ matching cycle in (b). Its two alternating
  perfect matchings yield the two terms of $I_L$, equivalently the two maximum
  independent current sets.  (c) Root-of-unity fusion forms the spectral-orbit
  matching cycle, and the equality shown closes the two cycles exactly.}
  \label{fig:prl-two-cycle-closure}
\end{figure}

\section{The finite spectral system and algebraic completeness}
\label{sec:finite-spectral-system}

We now turn this identity into a finite spectral problem.

\begin{definition}
\label{def:finite-spectral-equations}
With \(\lambda_{\mathrm o}:=\prod_{j=1}^{L}\lambda_{2j-1}\) and
\(\lambda_{\mathrm e}:=\prod_{j=1}^{L}\lambda_{2j}\), define the normalized
trial polynomial
\begin{equation}
\begin{aligned}
  \widetilde\Lambda_q(u;\boldsymbol c)
  ={}1+\sum_{k=1}^{L-1}c_k u^k+(-1)^L\bigl(\omega^q\lambda_{\mathrm o}
  +\lambda_{\mathrm e}\bigr)u^L.
\end{aligned}
  \label{eq:trial-transfer-polynomial}
\end{equation}
Here \(\boldsymbol c=(c_1,\dots,c_{L-1})^T\).  Cyclic covariance under
\(u\mapsto\omega u\), together with the fixed constant and leading
coefficients, gives the finite closure defect
\begin{equation}
  \Phi_q[\widetilde\Lambda_q](u)
  -\mathcal Z_{\boldsymbol\lambda}^{(N)}(u^N)
  =\sum_{\ell=1}^{L-1}
  R_{q,\ell}(\boldsymbol c)u^{N\ell}.
  \label{eq:finite-closure-defect}
\end{equation}
Thus, Eq.~\eqref{eq:matching-polynomial-closure} is equivalent to the common
vanishing of these coefficients.
\end{definition}

For completeness, put
\[
 a=(-1)^L(\omega^q\lambda_o+\lambda_e),\qquad
 b=\omega^q\lambda_{\mathrm{tot}},\qquad
 K=\begin{pmatrix}a&-b\\1&0\end{pmatrix}.
\]
The eigenvalues of \(K\) are \((-1)^L\omega^q\lambda_o\) and
\((-1)^L\lambda_e\). The highest-degree part of the ordered
spectral-cycle product consequently gives
\[
 [u^{NL}]\Phi_q[\widetilde\Lambda_q](u)
 =(-1)^{L(N-1)}\operatorname{tr}K^N
 =(-1)^L(\lambda_o^N+\lambda_e^N).
\]
This equals the contribution of the two perfect matchings to the highest
coefficient of \(Z^{(N)}_\lambda(u^N)\). The constant coefficients
are both one. Together with cyclic covariance, this leaves precisely
the \(L-1\) coefficients in \eqref{eq:finite-closure-defect}. 
The same calculation applies when the two eigenvalues of \(K\) coincide.

\begin{lemma}
\label{lem:trial-transfer-endpoints}
If \(\widetilde\Lambda_q(u;\boldsymbol c)\) is a physical transfer
eigenvalue in \(\mathcal H_q\) with Hamiltonian eigenvalue \(E\), then
\(c_1=-E\), and its leading coefficient is fixed by
\begin{equation*}
  I_L|_{\mathcal H_q}
  =\omega^q\lambda_{\mathrm o}+\lambda_{\mathrm e}.
\end{equation*}
\end{lemma}

\begin{proof}
The coefficient of \(-u\) in \(t(u)=I(u)\) is \(H\), 
while Eq. (\ref{eq:endpoint-matching-charges}) fixes the coefficient of \((-u)^L\) in sector \(q\).
\end{proof}

The root-of-unity closure thus yields the finite polynomial spectral system
\begin{equation}
  R_{q,\ell}(\boldsymbol c)=0,
  \qquad \ell=1,\ldots,L-1.
  \label{eq:finite-spectral-equations}
\end{equation}

\begin{proposition}
\label{prop:physical-finite-spectral-equations}
Assume that \(\lambda_{2j-1}\neq0\) for \(j=1,\ldots,L\).  Every physical
transfer eigenvalue in \(\mathcal H_q\) gives a solution of
Eq.~\eqref{eq:finite-spectral-equations}.
\end{proposition}

\begin{proof}
For every physical transfer eigenvalue, Theorem
\ref{thm:matching-polynomial-closure} makes the left-hand side of
Eq.~\eqref{eq:finite-closure-defect} vanish.  Hence all its coefficients
\(R_{q,\ell}(\boldsymbol c)\) vanish.
\end{proof}

Fix \(q\in\mathbb Z_N\), set \(n:=L-1\), and write the unknown coefficient
vector as \(\boldsymbol c=(c_1,\ldots,c_n)\in\mathbb C^n\).  All couplings
are regarded as fixed parameters and are suppressed from the notation.  The
finite spectral equations form the polynomial system
\begin{equation*}
  R_{q,1}(\boldsymbol c)
  =\cdots=
  R_{q,n}(\boldsymbol c)=0,
  \qquad
  R_{q,\ell}\in\mathbb C[c_1,\ldots,c_n].
\end{equation*}

For a polynomial in \(c_1,\ldots,c_n\), a superscript \([d]\) denotes the
sum of all monomials in which the powers of the variables \(c_j\) add up to
\(d\).

\begin{lemma}
\label{lem:finite-system-leading-forms}
For every \(\ell=1,\ldots,n\), the polynomial
\(R_{q,\ell}\) has total degree exactly \(N\), and its leading forms satisfy
\begin{equation}
  \sum_{\ell=1}^{n}R_{q,\ell}^{[N]}(\boldsymbol c)u^{N\ell}
  =\prod_{r=0}^{N-1}
  \left[
    \sum_{j=1}^{n}c_j(\omega^r u)^j
  \right].
  \label{eq:leading-form-product}
\end{equation}
\end{lemma}

\begin{proof}
By Eqs.~\eqref{eq:weighted-cycle-matching-trace} and
\eqref{eq:spectral-matching-polynomial}, a nonempty matching with
\(a\geq1\) edges leaves \(N-2a\leq N-2\) uncovered vertex weights.  Since
each vertex weight is affine-linear in \(\boldsymbol c\), whereas the edge
weights are independent of \(\boldsymbol c\), every nonempty-matching
contribution has total degree at most \(N-2\).  Only the empty matching can
contribute to total degree \(N\), and its contribution is
\begin{equation*}
  \prod_{r=0}^{N-1}
  \widetilde\Lambda_q(\omega^r u;\boldsymbol c).
\end{equation*}
The scalar term \(\mathcal Z_{\boldsymbol\lambda}^{(N)}(u^N)\) is independent
of \(\boldsymbol c\).  Consequently, \(\deg R_{q,\ell}\leq N\).  Keeping
only the middle summation term in
Eq.~\eqref{eq:trial-transfer-polynomial} in every factor gives
Eq.~\eqref{eq:leading-form-product}.  The product is invariant under
\(u\mapsto\omega u\), so it contains only powers of \(u^N\).
If \(c_\ell\) is the only nonzero coordinate, the right-hand side of
Eq.~\eqref{eq:leading-form-product} is
\begin{equation*}
  \omega^{\ell N(N-1)/2}c_\ell^N u^{N\ell},
\end{equation*}
which is nonzero.  Hence \(\deg R_{q,\ell}\geq N\). Together with the upper
bound above, this gives \(\deg R_{q,\ell}=N\).
\end{proof}

\begin{lemma}
\label{lem:leading-forms-common-zero}
The degree-\(N\) leading forms of the finite spectral equations have no
common zero other than the origin.
\end{lemma}

\begin{proof}
If \(R_{q,\ell}^{[N]}(\boldsymbol c)=0\) for all
\(\ell=1,\ldots,n\), the right-hand side of
Eq.~\eqref{eq:leading-form-product} is the zero polynomial in
\(\mathbb C[u]\).  Since \(\mathbb C[u]\) is an integral domain, i.e., it
has no zero divisors, a product can vanish identically only if at least one
of its factors vanishes identically.  The substitution
\(u\mapsto\omega^r u\) is invertible, so this implies
\begin{equation}
  \sum_{j=1}^{n}c_j u^j\equiv0,
  \qquad\text{and hence}\qquad
  c_1=\cdots=c_n=0.
  \label{eq:leading-forms-only-zero}
\end{equation}
\end{proof}

\begin{proposition}
\label{prop:no-solutions-at-infinity}
The homogenized finite spectral equations have no common zero at infinity.
\end{proposition}

\begin{proof}
Introduce the complex projective space \(\mathbb P^n(\mathbb C)\).  A point
is represented by a nonzero tuple
\([\boldsymbol c:w]=[c_1:\cdots:c_n:w]\).  These brackets denote homogeneous,
or projective, coordinates rather than a unique ordinary
\((n+1)\)-component vector. For every \(\lambda\in\mathbb C^\times\),
\begin{equation*}
  [c_1:\cdots:c_n:w]
  =[
    \lambda c_1:\cdots:\lambda c_n:\lambda w
  ].
\end{equation*}
The all-zero tuple is excluded,
\begin{equation*}
  (c_1,\ldots,c_n,w)\neq(0,\ldots,0).
\end{equation*}
Although there are \(n+1\) coordinates, the common rescaling removes one
independent degree of freedom, so the projective space has dimension \(n\).

When \(w\neq0\), choosing \(\lambda=w^{-1}\) normalizes the last coordinate
to one:
\begin{equation*}
  [c_1:\cdots:c_n:w]
  =\left[
    \frac{c_1}{w}:\cdots:\frac{c_n}{w}:1
  \right].
\end{equation*}
Thus the chart \(w\neq0\) is identified with the affine space
\(\mathbb C^n\). In particular,
\begin{equation*}
  [c_1:\cdots:c_n:1]
  \longleftrightarrow
  (c_1,\ldots,c_n)\in\mathbb C^n.
\end{equation*}
When \(w=0\), this normalization is impossible.  The points
\begin{equation*}
  [c_1:\cdots:c_n:0],
  \qquad
  (c_1,\ldots,c_n)\neq\boldsymbol0,
\end{equation*}
form the hyperplane at infinity.

Since \(\deg R_{q,\ell}=N\), homogenize each equation to total degree \(N\)
by
\begin{equation}
  \widehat R_{q,\ell}(\boldsymbol c,w)
  :=w^N R_{q,\ell}(\boldsymbol c/w)
  =\sum_{d=0}^{N}w^{N-d}R_{q,\ell}^{[d]}(\boldsymbol c).
  \label{eq:homogenized-finite-system}
\end{equation}
At \(w=1\), these are the original affine equations.  At \(w=0\), they
reduce to
\(\widehat R_{q,\ell}(\boldsymbol c,0)
=R_{q,\ell}^{[N]}(\boldsymbol c)\).  Consequently, a common zero at infinity
would have to satisfy
\begin{equation*}
  R_{q,1}^{[N]}(\boldsymbol c)
  =\cdots=
  R_{q,n}^{[N]}(\boldsymbol c)=0,
  \qquad
  \boldsymbol c\neq\boldsymbol0.
\end{equation*}
Equation~\eqref{eq:leading-forms-only-zero}, however, shows that these
equations force \(\boldsymbol c=\boldsymbol0\).  The all-zero tuple does not
define a projective point, so the homogenized equations have no common zero
at infinity.
\end{proof}

\begin{proposition}
\label{prop:finite-projective-zero-set}
The common projective zero set of the homogenized finite spectral equations
is zero-dimensional and therefore finite.
\end{proposition}

\begin{proof}
By the projective dimension theorem, any positive-dimensional projective
component either lies in the hyperplane \(w=0\) or intersects it
\cite{Hartshorne1977}.  Both
possibilities are excluded here.  Hence the common projective zero set is
zero-dimensional and therefore finite.
\end{proof}

\Needspace{15\baselineskip}
\begin{corollary}
\label{cor:bezout-finite-system}
The polynomial system in Eq.~\eqref{eq:finite-spectral-equations} has
\begin{equation}
  N^n
  =N^{L-1}
  =\dim\mathcal H_q
  \label{eq:bezout-finite-system-count}
\end{equation}
finite solutions, counted with multiplicity.
\end{corollary}

\begin{proof}
By B\'ezout's theorem, 
\begin{quote}
if the equations
\(F_0=\cdots=F_{n-1}=0\) have degrees \(d_0,\ldots,d_{n-1}\) and finitely
many solutions in \(\mathbb P^n\), then the number of solutions (counted with
multiplicity) is \(d_0\cdots d_{n-1}\)\cite{CLO2005}.
\end{quote}
Here \(F_0,\ldots,F_{n-1}\in
\mathbb C[x_0,\ldots,x_n]\) are homogeneous polynomials of positive degrees
\(d_0,\ldots,d_{n-1}\), and
\(V(F_0,\ldots,F_{n-1})\) denotes their set of common zeros in the complex
projective space \(\mathbb P^n\).  If \(m(p)\) is the multiplicity with which
a distinct common zero \(p\) enters the count, then the quoted statement is
equivalently
\begin{equation}
  \sum_{p\in V(F_0,\ldots,F_{n-1})}m(p)
  =\prod_{i=0}^{n-1}d_i.
  \label{eq:bezout-general}
\end{equation}
A simple isolated common zero has \(m(p)=1\). A degenerate common zero is
counted repeatedly through \(m(p)>1\).

All its hypotheses have now been verified.  There are \(n=L-1\) homogenized
equations \(\widehat R_{q,1},\ldots,\widehat R_{q,n}\). Each has degree \(N\). 
Their common projective zero set is zero-dimensional and hence finite, and
none of its points lies at infinity.  Therefore all of the projective
intersection multiplicity lies in the affine chart \(w=1\).  Applying
B\'ezout's theorem, the polynomial system has
\begin{equation*}
  \prod_{\ell=1}^{n}\deg\widehat R_{q,\ell}
  =N^n
  =N^{L-1}
  =\dim\mathcal H_q
\end{equation*}
solutions, counted with multiplicity.
\end{proof}

Corollary~\ref{cor:bezout-finite-system} 
fixes the total intersection multiplicity of the finite
spectral equations. We now identify this intersection multiplicity with
the algebraic multiplicity of the charge-sector energy spectrum.

\begin{proposition}[Energy resultant and algebraic completeness]
\label{prop:energy-resultant}
Let $n=L-1$, and let
$\widehat R_{q,1},\ldots,\widehat R_{q,n}$ be the degree-$N$
homogenizations defined in Eq.~(\ref{eq:homogenized-finite-system}). Define
\[
 \mathfrak R_q(E;\boldsymbol{\lambda})
 :=
 \operatorname{Res}_{c_1,\ldots,c_n,w}
 \left(
   \widehat R_{q,1},\ldots,\widehat R_{q,n},
   c_1+Ew
 \right).
\]
There is a nonzero constant $C_{N,L}$, depending only on $N$ and
$L$, such that
\[
 {\cal P}_q(E;\boldsymbol{\lambda})
 :=
 C_{N,L}^{-1}\mathfrak R_q(E;\boldsymbol{\lambda})
\]
is monic of degree $N^{L-1}$ in $E$ and satisfies
\[
 {\cal P}_q(E;\boldsymbol{\lambda})
 =
 \det_{{\cal H}_q}(E-H).
\]
Consequently, the finite spectral equations in Eq.~\eqref{eq:finite-spectral-equations} reproduce the
complete charge-sector energy spectrum with algebraic multiplicity.
The identity extends polynomially to arbitrary values of the physical
couplings.
\end{proposition}

\begin{proof}
Put $D=N^n=N^{L-1}$. By multihomogeneity of the homogeneous
resultant, $\mathfrak R_q$ has degree at most $D$ in the coefficients
of the final linear form $c_1+Ew$. Its coefficient of $E^D$ is,
up to the fixed convention for the resultant,
\[
 \operatorname{Res}
 \left(
   \widehat R_{q,1},\ldots,\widehat R_{q,n},w
 \right).
\]
This coefficient cannot vanish. Indeed, its vanishing would imply a
common projective zero of
$\widehat R_{q,1},\ldots,\widehat R_{q,n}$ with $w=0$, whereas
Proposition~\ref{prop:no-solutions-at-infinity} excludes such a zero. At $w=0$ the homogenized
equations reduce to their leading forms
$R^{[N]}_{q,\ell}(c)$, which are given by Eq.~\eqref{eq:leading-form-product} and are independent
of both $q$ and the physical couplings. Hence the coefficient of
$E^D$ is a nonzero constant $C_{N,L}$ depending only on $N$ and $L$.

The standard product formula for the multivariate resultant
\cite{CLO2005} therefore gives
\[
 {\cal P}_q(E;\boldsymbol{\lambda})
 =
 \prod_{p\in V_q(\boldsymbol{\lambda})}
 \bigl(E+c_1(p)\bigr)^{m(p)},
\]
where $V_q(\boldsymbol{\lambda})$ is the finite affine zero scheme of
Eq.~\eqref{eq:finite-spectral-equations} and $m(p)$ is the local intersection multiplicity. By
Lemma~\ref{lem:trial-transfer-endpoints}, $c_1=-E_{\rm phys}$ for a physical transfer eigenvalue.
Thus ${\cal P}_q$ is the monic energy-elimination polynomial of the
finite spectral equations.

It remains to identify it with the Hamiltonian characteristic
polynomial. To show that a simple-spectrum locus is nonempty, consider
the specialization
\[
 \lambda_{2j}=0,\qquad \lambda_{2j-1}=a_j\ne0 .
\]
Then
\[
 H=\sum_{j=1}^{L}a_j\tau_j .
\]
In the simultaneous $\tau_j$-eigenbasis, the states in ${\cal H}_q$
are labelled by
$\boldsymbol{s}=(s_1,\ldots,s_L)\in{\mathbb Z}_N^L$ with
$\sum_j s_j=q\pmod N$, and their energies are
\[
 E_{\boldsymbol{s}}
 =
 \sum_{j=1}^{L}a_j\omega^{s_j}.
\]
For two distinct configurations
$\boldsymbol{s}\ne\boldsymbol{s}'$, equality of the corresponding
energies requires
\[
 \sum_{j=1}^{L}
 a_j\bigl(\omega^{s_j}-\omega^{s'_j}\bigr)=0 .
\]
For each pair this is a proper linear condition on the parameters
$a_j$. Since only finitely many pairs occur, the $a_j$ may be chosen
so that none of these conditions is satisfied. Hence, for such a
choice, $H|_{{\cal H}_q}$ has exactly $D=N^{L-1}$ distinct
eigenvalues.

Simplicity of a finite-dimensional spectrum is stable under
sufficiently small perturbations of the matrix entries. We may
therefore perturb the even couplings from zero to sufficiently small
nonzero values while retaining a simple charge-sector spectrum.
Since the odd couplings are already nonzero, this gives a nonempty
open set of couplings contained in the chart on which
Proposition~\ref{prop:physical-finite-spectral-equations} applies.

On this open set every eigenspace of $H|_{{\cal H}_q}$ is
one-dimensional. Since the transfer family commutes with $H$, every
Hamiltonian eigenvector is then a common transfer-family eigenvector.
Proposition~\ref{prop:physical-finite-spectral-equations} 
consequently supplies $D$ solutions of Eq.~\eqref{eq:finite-spectral-equations} with
distinct values $c_1=-E$. Corollary~\ref{cor:bezout-finite-system} shows that the total
intersection multiplicity of all finite solutions is exactly $D$.
These physical solutions therefore exhaust the zero scheme and are
all simple. Hence
\[
 {\cal P}_q(E;\boldsymbol{\lambda})
 =
 \det_{{\cal H}_q}(E-H)
\]
throughout this open set.

Both sides are polynomials in $E$ and in the physical couplings.
Their difference therefore is a polynomial that vanishes on a
nonempty open subset of the coupling space, and hence vanishes
identically. The equality consequently extends to arbitrary values
of the physical couplings, including the singular coupling loci where
the particular cyclic-BBS parametrization used in
Theorem~\ref{thm:matching-polynomial-closure} is unavailable.
\end{proof}

Thus Eq.~\eqref{eq:finite-spectral-equations} gives not only the B\'ezout count of the finite
coefficient solutions, but the complete charge-sector energy spectrum
with algebraic multiplicity. At a nongeneric degeneracy or exceptional
point, the proposition fixes the algebraic multiplicity of the energy. 
It does not by itself determine the sizes of possible Jordan blocks.

Direct enumeration of the solutions of
Eq.~\eqref{eq:finite-spectral-equations}, like a direct use of the ODBA or
SoV spectral equations in Appendices
\ref{app:odba-finite-size-verification} and
\ref{app:separation-of-variables-solution}, respectively, becomes impractical
at large \(L\).

\section{The ground-state branch: Perron-Frobenius}
\label{sec:ground-state-branch}

The finite spectral equations contain every charge-sector transfer eigenvalue,
with \(c_1=-E\), but do not by themselves select the ground-state branch.  We
now prove that the
OBC and PBC ground states are the two endpoints of one eigenvalue branch.  Set
\(a_j:=\lambda_{2j-1}\) for \(j=1,\ldots,L\),
\(b_j:=\lambda_{2j}\) for \(j=1,\ldots,L-1\), and replace the closing
coupling \(\lambda_{2L}\) by a real seam parameter \(\mu\).  Thus
\begin{equation}
  H(\mu)
  =\sum_{j=1}^{L}a_j\tau_j
  +\sum_{j=1}^{L-1}b_j\sigma_j^\dagger\sigma_{j+1}
  +\mu\,\sigma_L^\dagger\sigma_1 .
  \label{eq:ground-seam-hamiltonian}
\end{equation}

\begin{assumption}
\label{ass:positive-reflection-couplings}
Throughout this section,
\begin{equation}
  a_j>0,\qquad b_j>0,\qquad \mu\geq0,
  \qquad
  a_j=a_{L+1-j},\qquad b_j=b_{L-j}.
  \label{eq:ground-reflection-conditions}
\end{equation}
The reflection identities are imposed for every index for which the
corresponding coupling is defined.
\end{assumption}

The homogeneous path \(a_j=1\), \(b_j=\lambda>0\), and
\(0\leq\mu\leq\lambda\) is a special case.  Since
Eq.~\eqref{eq:ground-seam-hamiltonian} uses the plus-sign convention, a ground
state in this section means an eigenstate whose energy has maximal real part.
Multiplication of the Hamiltonian by an overall minus sign converts this to
the usual lowest-real-part convention.  We write
\begin{equation*}
  Q:=\prod_{j=1}^{L}\tau_j=\omega^{\mathcal P}
\end{equation*}
for the global charge.

\begin{theorem}
\label{thm:perron-ground-state}
Under Assumption~\ref{ass:positive-reflection-couplings}, for every
\(N,L\geq2\) and \(\mu\geq0\), the Hamiltonian \(H(\mu)\) has a real,
algebraically simple eigenvalue \(E_*(\mu)\) such that
\begin{equation}
  \operatorname{Re}E<E_*(\mu)
  \quad\text{for every other eigenvalue }E.
  \label{eq:ground-spectral-bound}
\end{equation}
Its one-dimensional eigenspace belongs to the charge-zero sector:
\begin{equation}
  Q|\Psi_*(\mu)\rangle=|\Psi_*(\mu)\rangle .
  \label{eq:ground-charge-zero}
\end{equation}
\end{theorem}

The remainder of this section presents the proof of
Theorem~\ref{thm:perron-ground-state} and its consequences.
Together with the known OBC ground state, this theorem yields a unique analytic
continuation of the ground-state branch along the nonnegative seam path.  We
first establish the positivity properties needed for the proof.

We first record the finite-dimensional operator-algebraic setting.  For a
finite set \(S\), let
\begin{equation}
  \mathcal A:=\bigoplus_{s\in S}M_{d_s}(\mathbb C),
  \qquad
  \mathcal C:=\mathcal A_+
  =\bigoplus_{s\in S}M_{d_s}(\mathbb C)_+ .
  \label{eq:pf-cstar-algebra}
\end{equation}
With the operator norm, ordinary product, identity, and conjugate-transpose
involution, \(\mathcal A\) is a finite-dimensional \(C^*\)-algebra
\cite{EvansKawahigashi1998}.  Its self-adjoint part is denoted by
\(\mathcal A_{\mathrm h}\).  An element \(X=(X_s)_{s\in S}\) belongs to
\(\mathcal C\) when every block \(X_s\) is positive semidefinite, and it is
strictly positive, written \(X>0\), when every block is positive definite.
The latter elements form the relative interior of \(\mathcal C\) in
\(\mathcal A_{\mathrm h}\).

A complex-linear map \(\phi:\mathcal A\to\mathcal A\) is positive if it
preserves \(\mathcal C\), and strongly positive if
\begin{equation*}
  0\neq X\in\mathcal C\quad\Longrightarrow\quad \phi(X)>0.
\end{equation*}
It is completely positive if every amplification
\(\operatorname{id}_{M_n}\otimes\phi\) is positive.  A projection
\(p=p^*=p^2\) reduces \(\phi\) when
\(\phi(p\mathcal A p)\subseteq p\mathcal A p\). The map is irreducible if
only \(0\) and \(\boldsymbol1_{\mathcal A}\) reduce it.  A unital map is a
Schwarz map when
\begin{equation*}
  \phi(X^*X)\geq\phi(X)^*\phi(X)
  \qquad (X\in\mathcal A).
\end{equation*}
Every unital completely positive map is a Schwarz map.

We use two finite-dimensional Perron-Frobenius results.  A positive linear
map has a nonzero positive eigenvector at its spectral radius.  For an
irreducible Schwarz map, the peripheral spectrum is a finite subgroup of the
unit circle and each peripheral eigenvalue is algebraically simple. Moreover,
the peripheral spectrum is \(\{1\}\) precisely when every positive power of
the map is irreducible \cite[Thms.~2.5 and 4.2]{EvansHoeghKrohn1978}.
Here \(\operatorname{sp}(\phi)\) denotes the spectrum of \(\phi\), and
\(\mathbb T:=\{z\in\mathbb C:|z|=1\}\).

\begin{lemma}
\label{lem:positive-generator-criterion}
Let \(\mathcal G:\mathcal A\to\mathcal A\) have block form
\begin{equation}
  (\mathcal G X)_s
  =B_sX_s+X_sB_s^\dagger
  +\sum_{r\in S}\sum_{\alpha}
  V_{sr,\alpha}X_rV_{sr,\alpha}^\dagger .
  \label{eq:positive-generator}
\end{equation}
Suppose that the only family of subspaces
\((\mathcal W_s)_{s\in S}\) satisfying
\begin{equation}
  B_s\mathcal W_s\subseteq\mathcal W_s,
  \qquad
  V_{sr,\alpha}\mathcal W_r\subseteq\mathcal W_s
  \label{eq:positive-irreducibility}
\end{equation}
for every nonzero jump matrix is either the all-zero family or
\(\mathcal W_s=\mathbb C^{d_s}\) for every \(s\).  Then
\(e^{t\mathcal G}\) is completely positive and strongly positive for every
\(t>0\).
\end{lemma}

\begin{proof}
The first two terms of Eq.~\eqref{eq:positive-generator} generate the
congruences
\begin{equation*}
  X_s\longmapsto e^{tB_s}X_se^{tB_s^\dagger},
\end{equation*}
and each jump term is also a completely positive congruence.  The Duhamel
expansion of \(e^{t\mathcal G}\) is a norm-convergent sum of integrals of
compositions of these congruences.  Hence \(e^{t\mathcal G}\) is completely
positive and preserves \(\mathcal C\).

For \(0\neq X\in\mathcal C\), the support reached from \(X\) by the free
evolutions \(e^{uB_s}\) and the jumps \(V_{sr,\alpha}\) is the smallest
family satisfying Eq.~\eqref{eq:positive-irreducibility}.  In finite
dimension, invariance under every \(e^{uB_s}\) is equivalent to invariance
under \(B_s\).  Every finite jump path occurs on an open time simplex in the
Duhamel expansion.  If an output block had a nonzero null vector, positivity
would force every congruence along each such path to vanish on that vector.
Analyticity in the dwell times would then leave a proper reachable family,
contrary to the hypothesis.  Every output block therefore has full support,
so \(e^{t\mathcal G}X>0\).
\end{proof}

\Needspace{11\baselineskip}
\begin{lemma}
\label{lem:even-folding-irreducibility}
Under Assumption~\ref{ass:positive-reflection-couplings}, let \(L=2K\).
The Hamiltonian \(H(\mu)\) is linearly similar to the map
\begin{equation}
  \mathcal L_\mu(X)
  =BX+XB^\dagger
  +b_KA_KXA_K^\dagger
  +\mu A_1XA_1^\dagger
  \label{eq:ground-even-fold}
\end{equation}
on \(M_{N^K}(\mathbb C)\), where
\begin{equation*}
  B:=\sum_{j=1}^{K}a_j\tau_j
  +\sum_{j=1}^{K-1}b_j\sigma_j^\dagger\sigma_{j+1},
  \qquad
  A_K:=\sigma_K^\dagger,
  \qquad
  A_1:=\sigma_1.
\end{equation*}
The matrices in Eq.~\eqref{eq:ground-even-fold} satisfy the irreducibility
condition in Eq.~\eqref{eq:positive-irreducibility}.
\end{lemma}

\begin{proof}
Let
\(\mathcal H_{\mathrm L}:=\bigotimes_{j=1}^{K}\mathbb C_j^N\) and order the
right-half sites as \(2K,2K-1,\ldots,K+1\).  Reflection identifies the
ordered right space with the conjugate space
\(\overline{\mathcal H_{\mathrm L}}\).  In chosen bases, write a full-chain
vector as
\begin{equation}
  |X\rangle\!\rangle
  :=\sum_{r,s}X_{rs}|r\rangle\otimes\overline{|s\rangle}.
  \label{eq:ground-vectorization}
\end{equation}
This correspondence is complex-linear, and direct expansion gives
\begin{equation}
  (A\otimes\overline C)|X\rangle\!\rangle
  =|AXC^\dagger\rangle\!\rangle .
  \label{eq:ground-vectorization-identity}
\end{equation}
Here \(\overline C\,\overline{|s\rangle}:=\overline{C|s\rangle}\) on the
conjugate space.
The reflection conditions pair the two half-chain operators, while the middle
bond and the seam give the two congruence terms in
Eq.~\eqref{eq:ground-even-fold}.  This proves the stated similarity.  It is
used only as an algebraic positive-cone representation. No trace-preserving
physical dynamics is assigned to \(\mathcal L_\mu\).

It remains to exclude a proper common invariant subspace.  Since
\((\sigma_K^\dagger)^N=\boldsymbol1\), the algebra generated by \(B\) and
\(\sigma_K^\dagger\) also contains \(\sigma_K\).  The Weyl relation gives
\begin{equation}
  \sigma_K^\dagger B\sigma_K-B
  =a_K(\omega^{-1}-1)\tau_K ,
  \label{eq:ground-recover-tau}
\end{equation}
so this algebra contains \(\tau_K\).  If \(K>1\), the difference
\(\tau_KB\tau_K^\dagger-B\) is a nonzero multiple of
\(\sigma_{K-1}^\dagger\sigma_K\).  Multiplication by the already available
\(\sigma_K^\dagger\) recovers \(\sigma_{K-1}^\dagger\), and conjugation of
\(B\) by this operator recovers \(\tau_{K-1}\).  Iteration toward the left
recovers every local pair \(\sigma_j,\tau_j\). At each later step, a bond
already recovered at the preceding step is subtracted.

For a single clock, the matrices \(\sigma_j^r\tau_j^s\),
\(r,s\in\mathbb Z_N\), form a basis of \(M_N(\mathbb C)\).  Their tensor
products form a basis of \(M_{N^K}(\mathbb C)\).  A subspace invariant under
\(B\) and \(A_K\) is therefore invariant under every matrix and is either
zero or the whole space.  This proves the required irreducibility already at
\(\mu=0\). The seam jump is not needed.
\end{proof}

\begin{lemma}
\label{lem:odd-folding-irreducibility}
Under Assumption~\ref{ass:positive-reflection-couplings}, let \(L=2K+1\) and
\(c=K+1\).  After resolving the center site in the \(\sigma_c\)-basis, the
Hamiltonian is linearly similar to the map on
\(\bigoplus_{s\in\mathbb Z_N}M_{N^K}(\mathbb C)\) given by
\begin{equation}
  (\mathcal L_\mu X)_s
  =B_sX_s+X_sB_s^\dagger
  +\mu\sigma_1X_s\sigma_1^\dagger
  +a_cX_{s-1},
  \qquad s\in\mathbb Z_N,
  \label{eq:ground-odd-fold}
\end{equation}
where the center indices are read modulo \(N\) and
\begin{equation}
  B_s:=\sum_{j=1}^{K}a_j\tau_j
  +\sum_{j=1}^{K-1}b_j\sigma_j^\dagger\sigma_{j+1}
  +b_K\omega^s\sigma_K^\dagger .
  \label{eq:ground-odd-drift}
\end{equation}
The drift and jump matrices in Eq.~\eqref{eq:ground-odd-fold} satisfy the
irreducibility condition in Eq.~\eqref{eq:positive-irreducibility}.
\end{lemma}

\begin{proof}
The vectorization identity in Eq.~\eqref{eq:ground-vectorization-identity}
applied to the two reflected half chains gives
Eq.~\eqref{eq:ground-odd-fold}.  Its last term is a jump from block \(s-1\)
to block \(s\) with jump matrix \(\sqrt{a_c}\,\boldsymbol1\), so it connects
all blocks cyclically even at \(\mu=0\).

Let \((\mathcal W_s)\) satisfy
Eq.~\eqref{eq:positive-irreducibility}.  The center jump implies
\(\mathcal W_{s-1}\subseteq\mathcal W_s\) for every \(s\).  Going around the
finite cycle forces equality throughout, so all blocks share one subspace
\(\mathcal W\).  Drift invariance gives
\(B_s\mathcal W\subseteq\mathcal W\) for every \(s\).  For \(s\neq s'\),
\begin{equation}
  B_s-B_{s'}
  =b_K(\omega^s-\omega^{s'})\sigma_K^\dagger .
  \label{eq:ground-odd-difference}
\end{equation}
Hence the matrices \(B_s\) generate both \(\sigma_K^\dagger\) and the
left-half open-chain matrix obtained by subtracting the final term in
Eq.~\eqref{eq:ground-odd-drift}.  The site-by-site argument in the proof of
Lemma~\ref{lem:even-folding-irreducibility} again generates every local Weyl
pair and thus all of \(M_{N^K}(\mathbb C)\).  Therefore \(\mathcal W\) is
zero or the full left-half space, which proves the assertion.
\end{proof}

\begin{proposition}
\label{prop:ground-semigroup-strongly-positive}
Under Assumption~\ref{ass:positive-reflection-couplings}, for every
\(t>0\) and \(\mu\geq0\), the folded map
\begin{equation}
  \Phi_t:=e^{t\mathcal L_\mu}
  \label{eq:pf-semigroup-map}
\end{equation}
is completely positive and strongly positive on its finite-dimensional
\(C^*\)-algebra.
\end{proposition}

\begin{proof}
The even and odd folding maps have the form in
Eq.~\eqref{eq:positive-generator}.  Lemmas
\ref{lem:even-folding-irreducibility} and
\ref{lem:odd-folding-irreducibility} verify the corresponding invariant
subspace condition.  The conclusion follows from
Lemma~\ref{lem:positive-generator-criterion}.
\end{proof}

\begin{proof}[Proof of Theorem~\ref{thm:perron-ground-state}]
Fix \(t>0\), write \(r_t\) for the spectral radius of \(\Phi_t\), and use
Proposition~\ref{prop:ground-semigroup-strongly-positive}.  The positive-map
Perron-Frobenius theorem yields a nonzero \(R\in\mathcal C\) satisfying
\begin{equation}
  \Phi_t(R)=r_tR.
  \label{eq:pf-perron-element}
\end{equation}
Strong positivity forces \(R>0\) and \(r_t>0\).  Define the order
isomorphism \(T_R(X):=R^{1/2}XR^{1/2}\) and the normalized map
\begin{equation}
  \widetilde\Phi_t(X)
  :=r_t^{-1}T_R^{-1}\Phi_tT_R(X)
  =r_t^{-1}R^{-1/2}
    \Phi_t\!\left(R^{1/2}XR^{1/2}\right)R^{-1/2}.
  \label{eq:pf-doob-transform}
\end{equation}
The congruences \(T_R\) and \(T_R^{-1}\) are completely positive.
Consequently, \(\widetilde\Phi_t\) is completely positive, and
Eq.~\eqref{eq:pf-perron-element} gives
\(\widetilde\Phi_t(\boldsymbol1_{\mathcal A})
=\boldsymbol1_{\mathcal A}\).  It is therefore a Schwarz map.  It is also
strongly positive, and for every positive integer \(n\),
\begin{equation}
  \widetilde\Phi_t^n
  =r_t^{-n}T_R^{-1}\Phi_t^nT_R .
  \label{eq:pf-doob-powers}
\end{equation}
Every \(\Phi_t^n\) is strongly positive, so every
\(\widetilde\Phi_t^n\) is strongly positive and hence irreducible.  Indeed,
if a nonzero proper corner reduced one of these maps, the strongly positive
image of its supporting projection would have full support while remaining
in that proper corner, which is impossible.

The similarity and normalization give
\(r(\widetilde\Phi_t)=1\).  The stated Perron-Frobenius result for
irreducible Schwarz maps now implies
\begin{equation}
  \operatorname{sp}(\widetilde\Phi_t)\cap\mathbb T=\{1\},
  \qquad
  1\text{ is algebraically simple}.
  \label{eq:pf-normalized-peripheral-spectrum}
\end{equation}
Undoing the similarity and scalar normalization shows that \(r_t\) is an
algebraically simple eigenvalue of \(\Phi_t\), with eigenvector \(R>0\), and
that every other eigenvalue has modulus strictly less than \(r_t\).

Since \(\mathcal L_\mu\) commutes with \(\Phi_t\), the one-dimensional
Perron eigenspace is invariant under \(\mathcal L_\mu\), and hence
\begin{equation*}
  \mathcal L_\mu R=E_*(\mu)R .
\end{equation*}
The map \(\mathcal L_\mu\) preserves Hermiticity and \(R\) is positive
  definite, so \(E_*(\mu)\) is real and \(r_t=e^{tE_*(\mu)}\).  The
  algebraic simplicity of \(r_t\) excludes any other eigenvalue of
  \(\mathcal L_\mu\) from exponentiating to the same \(r_t\).  Spectral mapping
then gives
\begin{equation*}
  |e^{tE}|<e^{tE_*(\mu)}
  \quad\Longleftrightarrow\quad
  \operatorname{Re}E<E_*(\mu)
\end{equation*}
for every other eigenvalue \(E\).  A Jordan block at \(E_*(\mu)\) would
produce a Jordan block at the spectral radius of the exponential and is
therefore excluded.  Since the folding is a linear similarity, these
statements hold for the original Hamiltonian.

It remains to determine the charge.  For even \(L=2K\), put
\(U:=\prod_{j=1}^{K}\tau_j\). Under the folding, \(Q\) acts as the cone
automorphism \(X\mapsto UXU^\dagger\).  For odd \(L\), up to the choice of
the center-index shift direction, it acts as
\begin{equation*}
  (QX)_s=UX_{s-1}U^\dagger .
\end{equation*}
In both cases \(Q\) preserves the cone, \(Q^N=\boldsymbol1\), and
\([Q,\mathcal L_\mu]=0\).  Thus \(QR\) is another interior eigenvector at
the Perron eigenvalue.  Uniqueness of the interior ray gives \(QR=cR\) with
\(c>0\).  Since \(c^N=1\), one has \(c=1\), proving
Eq.~\eqref{eq:ground-charge-zero}.
\end{proof}

\begin{lemma}
\label{lem:obc-all-zero-ground-state}
At \(\mu=0\), the eigenvalue in Theorem~\ref{thm:perron-ground-state} is the
all-zero occupation energy of the open chain:
\begin{equation}
  E_*(0)=\sum_{k=1}^{L}\epsilon_k,
  \qquad \epsilon_k>0.
  \label{eq:obc-all-zero-energy}
\end{equation}
\end{lemma}

\begin{proof}
The exact free-parafermion OBC spectrum is~\cite{Fendley2014}
\begin{equation}
  E(\boldsymbol p)
  =\sum_{k=1}^{L}\epsilon_k\omega^{p_k},
  \qquad p_k\in\mathbb Z_N .
  \label{eq:ground-obc-spectrum}
\end{equation}
To determine the signs of the quasienergies, introduce
the lower-bidiagonal matrix
\begin{equation}
  \mathsf D_{jj}=a_j^{N/2},
  \qquad
  \mathsf D_{j+1,j}=b_j^{N/2},
  \label{eq:ground-obc-bidiagonal}
\end{equation}
whose remaining entries vanish.  The open-chain recurrence identifies
\(\epsilon_k^N\) with the eigenvalues of
\(\mathsf D^\dagger\mathsf D\), or equivalently
\(\epsilon_k=s_k(\mathsf D)^{2/N}\), where \(s_k\) denotes a singular
value and the positive real branch is used for the fractional power.  Every
diagonal entry of \(\mathsf D\) is nonzero, so \(\mathsf D\)
is invertible and every \(\epsilon_k\) is strictly positive.  Therefore
\begin{equation}
  \operatorname{Re}E(\boldsymbol p)
  =\sum_{k=1}^{L}\epsilon_k
  \cos\!\left(\frac{2\pi p_k}{N}\right)
  \leq\sum_{k=1}^{L}\epsilon_k,
  \label{eq:ground-obc-ordering}
\end{equation}
with equality only when \(p_1=\cdots=p_L=0\).  The unique Perron eigenvalue
at \(\mu=0\) is therefore the all-zero OBC occupation energy.
\end{proof}

\begin{corollary}
\label{cor:ground-state-analytic-continuation}
Under Assumption~\ref{ass:positive-reflection-couplings}, the all-zero OBC
ground energy and its one-dimensional eigenspace have a unique analytic
continuation along the entire positive seam path.  This continuation remains
the unique ground-state branch.  In the homogeneous case, its value at
\(\mu=\lambda\) is the PBC ground energy and belongs to
\(\mathcal H_0\).
\end{corollary}

\begin{proof}
The entries of \(H(\mu)\) are affine functions of \(\mu\), and
Theorem~\ref{thm:perron-ground-state} shows that \(E_*(\mu)\) is algebraically
simple for every \(\mu\geq0\).  Analytic perturbation theory therefore gives
a locally analytic eigenvalue and rank-one spectral projector near every
point of the seam path \cite[Chap.~II, Secs.~1--2]{Kato1995}.  Uniqueness
makes these local objects agree on overlaps, and the eigenvector may be chosen
locally analytically after fixing a normalization.  Lemma
\ref{lem:obc-all-zero-ground-state} identifies the branch at \(\mu=0\), while
the pointwise strict inequality in Eq.~\eqref{eq:ground-spectral-bound} shows
that the continuation remains the unique global max-real state.  At the
homogeneous endpoint \(\mu=\lambda\), it is therefore the PBC ground state.
Equation~\eqref{eq:ground-charge-zero} places it in \(\mathcal H_0\).
\end{proof}

\begin{remark}
The argument applies to odd and even \(N\).  For even \(N\), other OBC
patterns with \(p_k\in\{0,N/2\}\) give real energies, but
Eq.~\eqref{eq:ground-obc-ordering} places them strictly below the all-zero
state, and Theorem~\ref{thm:perron-ground-state} prevents them from becoming
the ground state along the seam.  Reflection symmetry is essential to the
folding proof above. It is not claimed to be necessary for the spectral
conclusion.  No conclusion is asserted here for nonpositive or complex
couplings, for a failure of reflection symmetry, or for selecting the
max-real state inside a prescribed nonzero charge sector.
\end{remark}

\section{Continuation of the physical branch}
\label{sec:physical-branch-continuation}

We now reopen the last bond while keeping the bulk couplings fixed.  For
\(\lambda>0\) and \(0\leq\mu\leq\lambda\), Eq.
\eqref{eq:ground-seam-hamiltonian} becomes
\begin{equation}
  H(\lambda,\mu)
  =\sum_{j=1}^{L}\tau_j
   +\lambda\sum_{j=1}^{L-1}\sigma_j^\dagger\sigma_{j+1}
   +\mu\,\sigma_L^\dagger\sigma_1.
  \label{eq:continuation-seam-hamiltonian}
\end{equation}
By Corollary~\ref{cor:ground-state-analytic-continuation}, the all-zero OBC
ground state has a unique analytic continuation to the PBC ground state, and
this branch lies in \(\mathcal H_0\).  Let \(P_*(\mu)\) be its rank-one
spectral projector.  Since the matching charges commute with the Hamiltonian,
their restrictions to \(\operatorname{Ran}P_*(\mu)\) are scalar.  Thus the
ground-state transfer eigenvalue may be written as
\begin{equation}
  \Lambda_*(u;\mu)
  =\widetilde\Lambda_0(u;\boldsymbol c_*(\mu)),
  \qquad
  c_{*,k}(\mu)
  =(-1)^k\operatorname{Tr}\!\left(P_*(\mu)I_k(\mu)\right),
  \label{eq:physical-transfer-coefficient-branch}
\end{equation}
for \(k=1,\ldots,L-1\).  These coefficients are analytic in \(\mu\), and
\(c_{*,1}(\mu)=-E_*(\mu)\).

At \(\mu=0\), the open-chain free-parafermion solution
\cite{Fendley2014,HenryBatchelor2023} gives
\begin{equation}
  P_L(u):=\Lambda_*(u;0)
  =\prod_{k=1}^{L}\bigl(1-\epsilon_{L,k}u\bigr),
  \qquad \epsilon_{L,k}>0.
  \label{eq:continuation-obc-ground-polynomial}
\end{equation}
The initial coefficient vector is therefore
\begin{equation}
  \boldsymbol c^{[0]}
  :=\bigl([u]P_L(u),\ldots,[u^{L-1}]P_L(u)\bigr)^T.
  \label{eq:continuation-initial-coefficients}
\end{equation}

\begin{definition}
\label{def:continuation-closure-jacobian}
Restore the seam dependence suppressed in Section
\ref{sec:finite-spectral-system} and set
\begin{equation}
\begin{aligned}
  \boldsymbol R_0(\boldsymbol c,\mu)
  &:={}
  \bigl(R_{0,1}(\boldsymbol c;\mu),\ldots,
        R_{0,L-1}(\boldsymbol c;\mu)\bigr)^T,\\
  \boldsymbol J_0(\boldsymbol c,\mu)
  &:={}
  D_{\boldsymbol c}\boldsymbol R_0(\boldsymbol c,\mu)
  =\left(
    \frac{\partial R_{0,\ell}}{\partial c_k}
   \right)_{\ell,k=1}^{L-1}.
\end{aligned}
  \label{eq:continuation-closure-and-jacobian}
\end{equation}
\end{definition}

Proposition~\ref{prop:physical-finite-spectral-equations} gives
\begin{equation}
  \boldsymbol R_0(\boldsymbol c_*(\mu),\mu)=\boldsymbol0,
  \qquad 0\leq\mu\leq\lambda.
  \label{eq:continuation-physical-root}
\end{equation}

\begin{proposition}[Regularity of the physical coefficient branch]
\label{prop:physical-jacobian}
Under Assumption~7.1, the physical solution
$c_*(\mu)$ of Eq.~(\ref{eq:continuation-physical-root}) is a simple zero of the finite spectral
system for every $0 \le \mu \le \lambda$. Consequently,
\[
 \det J_0(c_*(\mu),\mu) \ne 0,\qquad 0 \le \mu \le \lambda.
\]
\end{proposition}

\begin{proof}
Fix $0 \le \mu \le \lambda$. By Theorem~7.1, the ground-state energy
$E_*(\mu)$ is algebraically simple as an eigenvalue of $H(\mu)$.
By Proposition~6.4, the normalized energy resultant of the finite
spectral equations is the charge-sector characteristic polynomial.
Hence $E_*(\mu)$ is a simple zero of the energy resultant.

The physical transfer eigenvalue supplies at least one solution
$p_*(\mu)$ of the coefficient equations with
\[
 c_1(p_*(\mu))=-E_*(\mu).
\]
In the product representation of the normalized resultant from
Proposition~6.4, the multiplicity of the factor
$E-E_*(\mu)$ is the sum of the local intersection multiplicities of
all coefficient solutions having $c_1=-E_*(\mu)$. Since this
multiplicity is one, there is exactly one such solution and its local
intersection multiplicity is one.

For a zero-dimensional system of $L-1$ polynomial equations in
$L-1$ variables, an isolated zero has local intersection
multiplicity one precisely when it is a simple zero, equivalently
when the coefficient Jacobian has full rank. Therefore
\[
 \det J_0(c_*(\mu),\mu)\neq0.
\]
\end{proof}

Proposition 8.1 shows that the coefficient Jacobian is invertible at every point of the physical branch. The branch can therefore be followed directly through the coefficient equations without enumerating the remaining solutions of the finite spectral system.

\begin{lemma}
\label{lem:conditional-implicit-continuation}
Suppose that $R_0(c_0,\mu_0)=0$ and
$\det J_0(c_0,\mu_0)\neq0$. Then there is a unique local analytic
solution $c(\mu)$ through $(c_0,\mu_0)$, and
\begin{equation}
 \frac{dc}{d\mu}
 =
 -J_0(c,\mu)^{-1}
 \frac{\partial R_0}{\partial\mu}(c,\mu).
  \label{eq:conditional-jacobian-continuation}
\end{equation}
\end{lemma}

\begin{proof}
The map \(\boldsymbol R_0\) is polynomial in
\((\boldsymbol c,\mu)\).  The analytic implicit function theorem gives the
local solution.  Differentiating
\(\boldsymbol R_0(\boldsymbol c(\mu),\mu)=\boldsymbol0\) and multiplying by
\(\boldsymbol J_0^{-1}\) gives
Eq.~\eqref{eq:conditional-jacobian-continuation}.
\end{proof}

\begin{algorithm}
\label{alg:physical-branch-euler-continuation}
Choose a partition
\begin{equation*}
  0=\mu_0<\mu_1<\cdots<\mu_M=\lambda,
  \qquad
  \Delta\mu_m:=\mu_{m+1}-\mu_m,
\end{equation*}
and start with \(\boldsymbol c^{[0]}\) from
Eq.~\eqref{eq:continuation-initial-coefficients}.  Whenever the displayed
inverse exists, define
\begin{equation}
  \boldsymbol c^{[m+1]}
  =\boldsymbol c^{[m]}-\Delta\mu_m
  \left.
  \left(
    \boldsymbol J_0^{-1}
    \frac{\partial\boldsymbol R_0}{\partial\mu}
  \right)
  \right|_{\boldsymbol c=\boldsymbol c^{[m]},\,\mu=\mu_m}.
  \label{eq:physical-branch-euler-continuation}
\end{equation}
This is the explicit Euler discretization of
Eq.~\eqref{eq:conditional-jacobian-continuation}.  With a finite step size it gives
an approximation to the coefficient branch rather than an exact zero of the
closure system at the next parameter value. 
For the numerical calculations in Section 10, we use the usual predictor-corrector refinement of this continuation, with a secant predictor followed by damped Newton correction at the new coupling.
\end{algorithm}

\begin{corollary}
Under Assumption~7.1, the solution of Eq.~\eqref{eq:conditional-jacobian-continuation} issuing from the
initial coefficients in Eq.~\eqref{eq:continuation-initial-coefficients} is the physical coefficient branch
$c_*(\mu)$ on the whole interval $0\leq\mu\leq\lambda$.
If
\[
 h:=\max_m\Delta\mu_m,
\]
the Euler iterates in Eq.~\eqref{eq:physical-branch-euler-continuation} satisfy
\[
 \max_{0\leq m\leq M}
 \|c^{[m]}-c_*(\mu_m)\|=O(h),
 \qquad h\to0.
\]
Their endpoint therefore gives the PBC ground-state transfer
polynomial and energy,
\[
 E_{\rm PBC,L}(\lambda)
 =
 E_*(\lambda)
 =
 -c_{*,1}(\lambda)
 =
 -\lim_{h\to0}c^{[M]}_1 .
\]
\end{corollary}

\begin{proof}
Equation~\eqref{eq:physical-transfer-coefficient-branch} supplies an analytic
physical solution of Eq.~\eqref{eq:continuation-physical-root}.
By Proposition~8.1 its Jacobian is nonsingular at every point of the
seam path. Lemma~8.1 therefore makes this solution locally unique
throughout the interval, and the local solutions join to the unique
analytic branch selected in Corollary~7.1.
Since the physical branch is compact and
$\det J_0(c_*(\mu),\mu)\neq 0$ along it, Proposition~8.1 and
continuity give a neighbourhood of its graph on which $J_0^{-1}$
exists and the right-hand side of Eq.~\eqref{eq:conditional-jacobian-continuation} is analytic. It is
therefore locally Lipschitz in $c$, uniformly on the compact seam
interval. The Euler step in Eq.~\eqref{eq:physical-branch-euler-continuation} has local truncation error
$O((\Delta\mu_m)^2)$. The standard discrete Gronwall estimate then
gives
\[
\max_{0\leq m\leq M}
\bigl\|c^{[m]}-c_*(\mu_m)\bigr\|=O(h).
\]
For sufficiently small $h$, the iterates remain in this
neighbourhood, so the Jacobian inverses appearing in Eq.~\eqref{eq:physical-branch-euler-continuation} are
well defined. The endpoint identification then follows from
Corollary~7.1 and Lemma~\ref{lem:trial-transfer-endpoints}.
\end{proof}

\section{Asymptotics of the seam response}
\label{sec:seam-response-asymptotics}

Fix \(\lambda>0\) and consider the seam family in
Eq.~\eqref{eq:continuation-seam-hamiltonian}.  For brevity, write
\[
  \widetilde\Lambda_\mu(u):=\Lambda_*(u;\mu),
  \qquad
  c_k(\mu):=c_{*,k}(\mu),
  \qquad
  E(\mu):=E_*(\mu)=-c_1(\mu).
\]
Recall from Eq.~\eqref{eq:continuation-obc-ground-polynomial} that
\(P_L(u)=\widetilde\Lambda_0(u)\).

\begin{theorem}
\label{thm:finite-size-seam-response}
Assume that the zeros
\(r_j:=\epsilon_{L,j}^{-1}\), \(j=1,\ldots,L\), of \(P_L\) are simple and
that the contour rotations used below cross no additional poles.  Then the
finite-size seam response
\[
  E_0':=\left.\frac{\partial E(\mu)}{\partial\mu}\right|_{\mu=0}
\]
is represented by Eq.~\eqref{eq:seam-end-energy-coefficient}, with the
amplitudes given by the exact contour identity
\eqref{eq:seam-end-amplitudes-contour}, and equivalently by the real-axis
identities \eqref{eq:seam-end-energy-real-integral} and
\eqref{eq:seam-end-energy-reciprocal-integral} below.
\end{theorem}

\begin{proof}
Expanding Eq.~\eqref{eq:trial-transfer-polynomial} in powers of \(\mu\) gives
\begin{equation}
  \widetilde{\Lambda}_{\mu}(u)
  =P_L(u)+\mu\bigl[d_L(u)+(-1)^L\lambda^{L-1}u^L\bigr]
  +O(\mu^2),
  \label{eq:seam-mu-expansion}
\end{equation}
Writing \(\Phi_\mu(u):=\Phi_0[\widetilde{\Lambda}_\mu](u)\), differentiating
the closure identity in Eq.~\eqref{eq:finite-closure-defect} along the
physical root in Eq.~\eqref{eq:continuation-physical-root} at \(\mu=0\) gives
\begin{equation}
  \left.\frac{\mathrm d\Phi_\mu(u)}{\mathrm d\mu}\right|_{\mu=0}
  -\left.\frac{\mathrm d\mathcal Z_{\boldsymbol\lambda}^{(N)}(u^N)}
  {\mathrm d\mu}\right|_{\mu=0}=0.
  \label{eq:seam-differentiated-closure}
\end{equation}
By Eq.~\eqref{eq:classical-current-subset},
\(\mathcal Z_{\boldsymbol\lambda}^{(N)}(u^N)\) contains \(\mu\) only through
\(\mu^N\), so its derivative vanishes at \(\mu=0\).  By
Eqs.~\eqref{eq:spectral-matching-polynomial} and
\eqref{eq:weighted-cycle-matching-trace}, the edge weights of \(C_N\)
have no constant term on OBC and begin at first order in \(\mu\).  Therefore,
at \(\mu=0\), only the empty matching and the one-edge matchings contribute
to the derivative of \(\Phi_\mu(u)\).  For fixed \(u\), write
\(\Lambda_k(\mu):=\widetilde\Lambda_\mu(\omega^k u)\) and
\(\alpha_k(\mu):=\alpha_{0,k+1}(u)\), with the indices understood modulo
\(N\).  Then
\begin{equation}
\begin{gathered}
\begin{aligned}
  \left.\frac{\mathrm d\Phi_\mu(u)}{\mathrm d\mu}\right|_{\mu=0}
  ={}&\sum_{k=0}^{N-1}
  \left.\frac{\mathrm d\Lambda_k}{\mathrm d\mu}\right|_{\mu=0}
  \prod_{\substack{s=0\\s\ne k}}^{N-1}\Lambda_s(0)
  -\sum_{k=0}^{N-1}
  \left.\frac{\mathrm d\alpha_k}{\mathrm d\mu}\right|_{\mu=0}
  \prod_{\substack{s=0\\s\notin\{k,k+1\}}}^{N-1}\Lambda_s(0)
\end{aligned}
\\[-0.25ex]
\begin{aligned}
  ={}&\displaystyle\sum_{k=0}^{N-1}
  \bigl[d_L(\omega^k u)+(-1)^L \lambda^{L-1}(\omega^k u)^L\bigr]
  \prod_{\substack{s=0\\s\ne k}}^{N-1}P_L(\omega^s u)
  \\
  &\displaystyle{}-\lambda^{L-1}u^{2L}\sum_{k=0}^{N-1}\omega^{L+2Lk}
  \prod_{\substack{s=0\\s\notin\{k,k+1\}}}^{N-1}P_L(\omega^s u).
\end{aligned}
\end{gathered}
  \label{eq:seam-matching-derivative}
\end{equation}
At \(u=r_j\), one has \(P_L(r_j)=0\), and
Eq.~\eqref{eq:seam-matching-derivative} gives
\begin{equation}
\begin{aligned}
  d_L(r_j)={}&(-1)^{L-1}\lambda^{L-1}r_j^L
  +\lambda^{L-1}r_j^{2L}\left[
  \frac{\omega^L}{P_L(\omega r_j)}
  +\frac{\omega^{-L}}{P_L(\omega^{-1}r_j)}
  \right].
\end{aligned}
  \label{eq:seam-dL-at-obc-root}
\end{equation}
Thus, the Lagrange interpolating polynomial gives
\begin{equation}
\begin{aligned}
  d_L(u)
  =\sum_{j=1}^{L}
  \frac{d_L(r_j)P_L(u)}{(u-r_j)P_L'(r_j)}
  =\sum_{k=1}^{L-1}
  \left.\frac{\mathrm dc_k}{\mathrm d\mu}\right|_{\mu=0}u^k.
\end{aligned}
  \label{eq:seam-dL-lagrange-interpolation}
\end{equation}
Hence the coefficient linear in \(\mu\) in the Taylor expansion of the
ground-state branch energy about \(\mu=0\) is
\begin{equation}
\begin{aligned}
  \left.\frac{\partial E(\mu)}{\partial\mu}\right|_{\mu=0}
  =-\left.\frac{\mathrm dc_1}{\mathrm d\mu}\right|_{\mu=0}
  =-\left.\frac{d_L(u)}{u}\right|_{u=0}
  =\sum_{j=1}^{L}\frac{d_L(r_j)}{P_L'(r_j)r_j^2}.
\end{aligned}
  \label{eq:seam-ground-energy-linear-coefficient}
\end{equation}
The lowest power of \(u\) in \(d_L(u)\) is \(u^1\), so dividing by
\(u\) before setting \(u=0\) produces no singularity.
For \(0\leq k\leq L-2\), the function \(z^k/P_L(z)\) is
\(O(z^{-2})\) as \(z\to\infty\), so its residue at infinity vanishes.
The residue theorem therefore gives
\begin{equation}
  \sum_{j=1}^{L}\frac{r_j^k}{P_L'(r_j)}=0,
  \qquad 0\leq k\leq L-2.
  \label{eq:seam-residue-sum-rule}
\end{equation}
Using Eq.~\eqref{eq:seam-ground-energy-linear-coefficient} and eliminating the
first term in Eq.~\eqref{eq:seam-dL-at-obc-root} with
Eq.~\eqref{eq:seam-residue-sum-rule}, we obtain
\begin{equation}
\begin{aligned}
  E_0'
  =\omega^L A_{+,L}+\omega^{-L}A_{-,L}.
\end{aligned}
  \label{eq:seam-end-energy-coefficient}
\end{equation}
Here
\begin{equation}
  A_{\pm,L}
  :=\lambda^{L-1}\sum_{j=1}^{L}
  \frac{r_j^{2L-2}}{P_L'(r_j)P_L(\omega^{\pm1}r_j)}.
  \label{eq:seam-end-amplitudes}
\end{equation}
Equivalently, the residue theorem gives the exact finite-size contour identity
\begin{equation}
  A_{\pm,L}
  =\frac{\lambda^{L-1}}{2\pi i}
  \oint_{\Gamma}
  \frac{z^{2L-2}}{P_L(z)P_L(\omega^{\pm1}z)}\,\mathrm dz.
  \label{eq:seam-end-amplitudes-contour}
\end{equation}
Here \(\Gamma\) is positively oriented and encloses only the roots of
\(P_L(z)\) and no other poles.  The integrands in
Eq.~\eqref{eq:seam-end-amplitudes-contour} are \(O(z^{-2})\) at infinity, so
the arcs introduced in the following contour rotations have vanishing
contributions.
Make the change of variables \(s=e^{i\theta}z\), with
\(\theta=\pi/N\), so that \(\omega=e^{2i\theta}\).  For real \(s\),
\(P_L(z)P_L(\omega z)=|P_L(e^{-i\theta}s)|^2\).  Combining the two
signs in Eq.~\eqref{eq:seam-end-amplitudes-contour} and rotating the contours
to the real axis gives
\begin{equation}
\begin{aligned}
  E_0'
  =\frac{\sin\theta}{\pi}\lambda^{L-1}
  \int_{-\infty}^{\infty}
  \frac{s^{2L-2}\,\mathrm ds}
  {\prod_{j=1}^{L}
  \left[1-2s\epsilon_{L,j}\cos\theta+s^2\epsilon_{L,j}^2\right]}.
\end{aligned}
  \label{eq:seam-end-energy-real-integral}
\end{equation}
Define the exact finite-size factors
\(D_{N,\lambda;j,L}(x):=x^2-2x\epsilon_{L,j}\cos\theta
+\epsilon_{L,j}^2\).  Making the further change of variables \(x=1/s\)
gives
\begin{equation}
  E_0'
  =\frac{\sin\theta}{\pi}\lambda^{L-1}
  \int_{-\infty}^{\infty}
  \frac{\mathrm dx}
  {\prod_{j=1}^{L}D_{N,\lambda;j,L}(x)}.
  \label{eq:seam-end-energy-reciprocal-integral}
\end{equation}
\end{proof}
For the asymptotic analysis, fix \(0<\lambda<1\).  The OBC single-particle
energies have the form
\[
  \epsilon_{N,\lambda}(k)
  =\left(1+\lambda^N+2\lambda^{N/2}\cos k\right)^{1/N}
\]
with \(\epsilon_{L,j}=\epsilon_{N,\lambda}(k_{j,L})\).  All \(L\) momenta
lie in \((0,\pi)\) and satisfy
\[
  \sin((L+1)k_{j,L})=-\lambda^{N/2}\sin(Lk_{j,L})
\]
\cite{Fendley2014,AlcarazBatchelorLiu2017}.
Writing \(a=\lambda^{N/2}<1\), the quantization condition is equivalently
\[
  (L+1)k_{j,L}=j\pi+\arctan\!\frac{a\sin k_{j,L}}{1+a\cos k_{j,L}},
  \qquad j=1,\ldots,L.
\]
In particular,
\(0\leq k_{j,L}-j\pi/(L+1)<\pi/[2(L+1)]\), and the finite-size and
continuum energies obey the bounds
\[
  0<\epsilon_-:=(1-a)^{2/N}
  \leq\epsilon_{L,j},\epsilon_{N,\lambda}(k)
  \leq(1+a)^{2/N}=: \epsilon_+.
\]
Define the continuum factor
\[
  D_{N,\lambda}(x,k)
  :=x^2-2x\epsilon_{N,\lambda}(k)\cos\theta
  +\epsilon_{N,\lambda}(k)^2.
\]
For any continuous function \(f\) on \([0,\pi]\), the corresponding mode sum
satisfies
\begin{equation}
  \frac{1}{L}\sum_{j=1}^{L}f(k_{j,L})
  \longrightarrow\frac{1}{\pi}\int_{0}^{\pi}f(k)\,\mathrm dk.
  \label{eq:seam-obc-mode-riemann-sum}
\end{equation}
Consequently,
\begin{equation}
\begin{gathered}
  \frac{1}{L}\sum_{j=1}^{L}
  \ln D_{N,\lambda;j,L}(x)
  \longrightarrow \frac{1}{\pi}\int_{0}^{\pi}
  \ln D_{N,\lambda}(x,k)\,\mathrm dk,\\
  h_{N,\lambda}(x):=\frac{1}{\pi}\int_{0}^{\pi}
  \ln D_{N,\lambda}(x,k)\,\mathrm dk.
\end{gathered}
  \label{eq:seam-obc-log-denominator-limit}
\end{equation}
Define
\(\Delta_{N,\lambda}(x):=h_{N,\lambda}(x)-\ln\lambda\).
The error needed below is bounded, rather than required to tend to zero.
Indeed, \(D_{N,\lambda}(x,k)=|x-e^{i\theta}\epsilon_{N,\lambda}(k)|^2\)
is strictly positive, and
\[
  \sup_{x\in\mathbb R,\,k\in[0,\pi]}
  \left|\partial_k\ln D_{N,\lambda}(x,k)\right|
  \leq\frac{4a}{N(1-a)^2\sin\theta}.
\]
The preceding bound on the momenta and the Riemann-sum error therefore give
\[
  r_L(x):=\sum_{j=1}^{L}\ln D_{N,\lambda;j,L}(x)-Lh_{N,\lambda}(x),
  \qquad
  \sup_{x\in\mathbb R}|r_L(x)|\leq C_{N,\lambda},
\]
where \(C_{N,\lambda}\) is independent of \(L\).  In estimating the
Riemann-sum error, an \(x\)-dependent constant can be subtracted from the
integrand, since the normalized sum and integral have the same total
weight.  Equation~\eqref{eq:seam-end-energy-reciprocal-integral} thus reads
\begin{equation}
  E_0'
  =\frac{\sin(\pi/N)}{\pi\lambda}
  \int_{-\infty}^{\infty}
  e^{-L\Delta_{N,\lambda}(x)-r_L(x)}\,\mathrm dx.
  \label{eq:seam-end-asymptotic-integral}
\end{equation}


Define
\[
  \gamma_N(\lambda)
  :=\min_{x\in\mathbb R}\Delta_{N,\lambda}(x).
\]
This minimum is attained: \(h_{N,\lambda}\) is continuous and
\(h_{N,\lambda}(x)\geq2\ln(|x|-\epsilon_+)\) for
\(|x|>\epsilon_+\).  For \(R>\epsilon_+\), the same bound gives
\[
  \int_{|x|>R}e^{-Lh_{N,\lambda}(x)}\,\mathrm dx
  \leq\frac{2(R-\epsilon_+)^{1-2L}}{2L-1}.
\]
Choose \(R\) large enough that
\(2\ln(R-\epsilon_+)>\min_x h_{N,\lambda}(x)\).
On \([-R,R]\), the integral is at most
\(2R\exp[-L\min_x h_{N,\lambda}(x)]\).  Conversely, continuity gives a
neighborhood of a minimizer on which
\(h_{N,\lambda}(x)<\min_x h_{N,\lambda}(x)+\eta\) for any \(\eta>0\),
providing the corresponding exponential lower bound.  The bounded factor
\(e^{-r_L(x)}\) does not change these exponential rates.  Hence
\begin{equation}
  \lim_{L\to\infty}\frac{1}{L}\ln E_0'
  =-\gamma_N(\lambda),
  \qquad 0<\lambda<1.
  \label{eq:seam-end-steepest-descent}
\end{equation}
Thus \(\gamma_N>0\) gives exponential decay and \(\gamma_N<0\) gives
exponential growth.  A bounded correction to the logarithmic mode sum can
alter a prefactor but cannot move a zero of \(\gamma_N\).  No claim about
the prefactor or the behavior at zero exponential rate is needed for the
numerical search.

At a zero of \(\gamma_N\) attained at a nondegenerate global minimizer
\(x_*\), the equations to be solved are
\begin{equation}
\begin{aligned}
  \Delta_{N,\lambda}'(x_*)&=0,\\
  \Delta_{N,\lambda}(x_*)&=0,
\end{aligned}
  \label{eq:seam-critical-saddle-conditions}
\end{equation}
with \(\Delta_{N,\lambda}''(x_*)>0\).  Selecting the global minimum is
essential. A stationary point alone does not determine \(\gamma_N\).
If this global minimizer is unique and
\(\partial_\lambda\Delta_{N,\lambda}(x_*)\ne0\), the
implicit function theorem gives a smooth local minimizing branch and
\(\gamma_N'(\lambda)=\partial_\lambda\Delta_{N,\lambda}(x_*)\).
The zero is then a change of sign of the exponential rate.  The Jacobian
of the two equations, with respect to \((x,\lambda)\), has determinant
\(\partial_x^2\Delta_{N,\lambda}(x_*)
\partial_\lambda\Delta_{N,\lambda}(x_*)\ne0\), so Newton iteration
converges locally for a sufficiently close initial value.

The bounds used here are uniform when \(\lambda\) ranges over a compact
subinterval of \((0,1)\), but do not cover \(\lambda=1\).
For \(\lambda>1\), the OBC spectrum contains an exponentially small
quasienergy at large \(L\)~\cite{AlcarazBatchelorLiu2017}. The bounded-error
argument above does not apply.  The duality relation used in the next
section concerns the continuum exponent and the reciprocal candidate
critical point, not an extension of
Eq.~\eqref{eq:seam-end-steepest-descent} to the full seam response at
\(\lambda>1\).

\begin{lemma}
\label{lem:conditional-seam-three-halves-law}
Suppose that, near a solution \((\lambda_*,x_*)\) of
Eq.~\eqref{eq:seam-critical-saddle-conditions}, the exponent has the local
quadratic form \eqref{eq:seam-critical-parabola}.  Then the positive area of
the one-sided negative lobe of this approximating parabola and its second
derivative are given by
Eqs.~\eqref{eq:seam-critical-area} and
\eqref{eq:seam-critical-area-curvature}.  Consequently, if the
boundary-induced ground-state energy correction has the form
\(g(\lambda)\rho(\lambda)+R(\lambda)\), where \(g\) is smooth,
\(g(\lambda_*)\ne0\), and \(R\) is twice continuously differentiable near
\(\lambda_*\), its second derivative is singular at \(\lambda_*\).
\end{lemma}

\begin{proof}
Near \((\lambda_*,x_*)\), the exponent may be approximated by the parabola
\begin{equation}
  \Delta_{N,\lambda}(x)
  \sim-a(\lambda-\lambda_*)+b(x-x_*)^2,
  \qquad a,b>0.
  \label{eq:seam-critical-parabola}
\end{equation}
For \(\lambda>\lambda_*\), only the part of this parabola below zero
contributes.  
For the one-sided lobe area \(\rho\), only the region where the quadratic approximation is negative is included.
The positive area of the negative lobe of the approximating
parabola is
\begin{equation}
  \rho(\lambda)=\frac{4a^{3/2}}{3\sqrt b}
  (\lambda-\lambda_*)^{3/2}.
  \label{eq:seam-critical-area}
\end{equation}
Its second derivative is
\begin{equation}
  \rho''(\lambda)=\frac{a^{3/2}}{\sqrt b}
  (\lambda-\lambda_*)^{-1/2}.
  \label{eq:seam-critical-area-curvature}
\end{equation}
Extend \(\rho\) by zero for \(\lambda\leq\lambda_*\).
Thus \(\rho(\lambda)\) and its first derivative remain continuous at
\(\lambda_*\), while the second derivative is singular.  Under the additional
hypothesis on the ground-state energy correction in the statement, the term
\(g(\lambda)\rho(\lambda)\) has the same divergent leading second derivative,
whereas \(R''\) remains finite.  Its second derivative is therefore singular
at \(\lambda_*\).
\end{proof}

\begin{remark}
Equation~\eqref{eq:seam-end-steepest-descent} determines the exponential
rate of the ground-state response to the closing bond at \(\mu=0\).
We use a sign-changing zero of \(\gamma_N\) to propose a candidate
critical coupling for the periodic chain, and examine this prediction
using the finite-size PBC ground-state curvature in
Section~\ref{sec:homogeneous-n3-application}.
The interpretation of the area term in
Lemma~\ref{lem:conditional-seam-three-halves-law} as a singular
contribution to the thermodynamic PBC ground-state energy remains
conditional on the additional hypothesis stated there. 
The auxiliary variable $x$ entering the seam-response exponent is 
not identified here with physical lattice momentum.
\end{remark}

\section{Application to the homogeneous \texorpdfstring{\(Z_N\)}{ZN} chain}
\label{sec:homogeneous-n3-application}

We first consider the homogeneous chain for an arbitrary integer \(N\geq2\).
The quantities entering Eq.~\eqref{eq:seam-critical-saddle-conditions} are
\begin{equation}
\begin{aligned}
  \epsilon_{N,\lambda}(k)
  &=\left(1+\lambda^N+2\lambda^{N/2}\cos k\right)^{1/N},\\
  \Delta_{N,\lambda}(x)
  &=\frac{1}{\pi}\int_0^\pi
  \ln D_{N,\lambda}(x,k)\,\mathrm dk-\ln\lambda,
\end{aligned}
  \label{eq:application-dispersion-exponent}
\end{equation}
where
\begin{equation}
  D_{N,\lambda}(x,k)
  =x^2-2x\epsilon_{N,\lambda}(k)\cos(\pi/N)
  +\epsilon_{N,\lambda}(k)^2.
  \label{eq:application-denominator}
\end{equation}
\begin{lemma}[Finite-size duality in the neutral sector]
\label{lem:finite-duality}
For the homogeneous periodic chain
\[
H_{\rm PBC,L}(\lambda)
=
\sum_{j=1}^{L}\tau_j
+
\lambda\sum_{j=1}^{L}\sigma_j^\dagger\sigma_{j+1},
\qquad \sigma_{L+1}=\sigma_1,
\]
Kramers-Wannier duality exchanges the onsite and 
bond operators~\cite{FradkinSusskind1978,CobaneraOrtizNussinov2011}.
On the neutral untwisted sector ${\cal H}_0$, it maps the periodic
family to itself and gives the spectral equivalence
\[
\operatorname{spec}
\left(
H_{\rm PBC,L}(\lambda)\big|_{{\cal H}_0}
\right)
=
\lambda\,
\operatorname{spec}
\left(
H_{\rm PBC,L}(\lambda^{-1})\big|_{{\cal H}_0}
\right).
\]
In particular, the ground-state branch selected in Theorem~7.1
satisfies
\[
E_{\rm PBC,L}(\lambda)
=
\lambda E_{\rm PBC,L}(\lambda^{-1}).
\]
The same relation holds for the physical Hamiltonian
$H_{\rm phys}=-H_{\rm PBC}$.
\end{lemma}

\begin{proof}
Introduce dual Weyl variables on the bonds of the original chain by
\[
\widetilde{\tau}_{j+1/2}
=
\sigma_j^\dagger\sigma_{j+1},
\qquad
\widetilde{\sigma}_{j+1/2}
=
\prod_{k=1}^{j}\tau_k .
\]
They obey the same local Weyl relations as the original clock
variables, and
\[
\tau_j
=
\widetilde{\sigma}_{j-1/2}^{\dagger}
\widetilde{\sigma}_{j+1/2}.
\]
The dual boundary condition is determined by the original global
charge:
\[
\widetilde{\sigma}_{L+1/2}
=
Q\,\widetilde{\sigma}_{1/2},
\qquad
Q=\prod_{j=1}^{L}\tau_j .
\]
Thus charge and boundary twist are exchanged by duality~\cite{CobaneraOrtizNussinov2011}.  
In the neutral sector $Q=1$, the dual variables are periodic.  Moreover,
\[
\prod_{j=1}^{L}\widetilde{\tau}_{j+1/2}
=
\prod_{j=1}^{L}\sigma_j^\dagger\sigma_{j+1}
=1. 
\]
%
The dual variables therefore act on the \(N^{L-1}\)-dimensional neutral dual sector, 
with the redundant overall dual clock removed. 
The inverse construction exchanges the original and dual variables. 
Thus the transformation defines an isomorphism of the two neutral-sector operator algebras.
Therefore
\[
H_{\rm PBC,L}(\lambda)
\longmapsto
\sum_{j=1}^{L}
\widetilde{\sigma}_{j-1/2}^{\dagger}
\widetilde{\sigma}_{j+1/2}
+
\lambda\sum_{j=1}^{L}\widetilde{\tau}_{j+1/2}
=
\lambda H_{\rm PBC,L}(\lambda^{-1}),
\]
up to the relabelling of the dual lattice.  This proves the
charge-zero spectral relation.  Theorem~7.1 places the physical
ground-state branch in ${\cal H}_0$, giving the stated ground-energy
identity.
\end{proof}

For \(0<\lambda<1\), the sign of
\(\gamma_N(\lambda)=\min_x\Delta_{N,\lambda}(x)\) determines the
exponential rate of the OBC seam response.  A change of sign can therefore
be located by minimizing \(\Delta_{N,\lambda}\) at fixed \(\lambda\) and
then finding a zero of the minimum, whenever such a zero exists.
The numerical search for solutions of
Eq.~\eqref{eq:seam-critical-saddle-conditions} is restricted to the coupling
interval \([0,1]\).  Only \(0<\lambda\leq1\) is sampled, with \(\lambda=0\)
understood as a limiting endpoint because \(\ln\lambda\) is undefined
there.  The corresponding results for \(\lambda\in[1,\infty)\) are
obtained by duality from those at \(\lambda^{-1}\in(0,1]\), using 
the continuum duality 
\begin{equation}
  \epsilon_{N,\lambda^{-1}}(k)=\lambda^{-1}\epsilon_{N,\lambda}(k),
  \qquad
  \Delta_{N,\lambda^{-1}}(x)=\Delta_{N,\lambda}(\lambda x),
  \label{eq:application-duality}
\end{equation}
and hence \(\gamma_N(\lambda^{-1})=\gamma_N(\lambda)\).  No independent
numerical scan is performed for \(\lambda>1\).  The endpoint \(\lambda=1\)
is evaluated separately from the continuum integral. Differentiability at
its minimizing point is not assumed.  The exponential-rate argument is
applied only to roots lying strictly inside \((0,1)\).

For fixed \(N\), we first perform a coarse scan of the coupling interval
\([0,1]\) with step size \(0.001\).  At each sampled value
\(0<\lambda<1\), we evaluate the integrals by numerical quadrature and
minimize \(\Delta_{N,\lambda}(x)\) over \(x\in[0,2]\), comparing its values
at local minima and at the endpoints.  The stationary-point equation is
\begin{equation}
  \partial_x\Delta_{N,\lambda}(x)
  =\frac{1}{\pi}\int_0^\pi
  \frac{2x-2\epsilon_{N,\lambda}(k)\cos(\pi/N)}
  {D_{N,\lambda}(x,k)}\,\mathrm dk
  =0.
  \label{eq:application-stationary-equation}
\end{equation}
For \(0<\lambda<1\), the derivative is negative for \(x<0\) and positive for
\(x>\cos(\pi/N)(1+\lambda^{N/2})^{2/N}\).  Consequently, every global
minimizer lies in
\begin{equation}
  \left[0,\,\cos(\pi/N)(1+\lambda^{N/2})^{2/N}\right]\subset[0,2].
  \label{eq:application-minimizer-interval}
\end{equation}
For \(N=2\), one has \(\cos(\pi/2)=0\), so the unique minimizer is
\(x_*=0\).  The continuum integral can be evaluated explicitly:
\begin{equation}
\begin{aligned}
  \gamma_2(\lambda)
  &=\Delta_{2,\lambda}(0)
   =\frac1\pi\int_0^\pi
     \ln(1+\lambda^2+2\lambda\cos k)\,\mathrm dk-\ln\lambda\\
  &=\ln\frac{1+\lambda^2+|1-\lambda^2|}{2\lambda}
   =|\ln\lambda|.
\end{aligned}
  \label{eq:application-ising-exponent}
\end{equation}
Thus the only zero is \(\lambda_c=1\), which is fixed by the duality
transformation. There are no two distinct reciprocal zeros.  This agrees
with the critical coupling of the transverse-field Ising chain
\cite{AlcarazBatchelorLiu2017}.  Since \(\gamma_2(\lambda)\geq0\) on both
sides, this zero is not detected by a change of sign and must instead be
checked at the endpoint of \([0,1]\).  Moreover,
\begin{equation}
  \Delta_{2,1}(x)
  =2\operatorname{arsinh}\!\left(\frac{|x|}{2}\right)
  =|x|+O(|x|^3).
  \label{eq:application-ising-endpoint}
\end{equation}
The function is not differentiable at its minimizer \(x=0\).  The smooth
nondegenerate stationary-point assumptions accompanying
Eq.~\eqref{eq:seam-critical-saddle-conditions}, and the associated Newton
argument, therefore do not apply at this endpoint.  Its zero is obtained
directly from the continuum integral.

For interior zeros, we next compare
\(\gamma_N(\lambda)=\min_x\Delta_{N,\lambda}(x)\) at
successive coupling values.  Opposite signs at the endpoints of a small
interval identify where to perform high-precision root finding.
Global minimization and the endpoint signs are then checked for the chosen
\(N\).

For example, for \(N=3\), this coarse scan detects opposite signs at the
endpoints of \([0.784,0.785]\), with
\begin{equation}
  \gamma_3(0.784)=0.000650625756030\ldots,
  \qquad
  \gamma_3(0.785)=-0.000401056569291\ldots,
  \label{eq:application-n3-sign-bracket}
\end{equation}

Within this bracket, uniqueness of the global minimizer can be established
analytically.  To see this, put
\(A=(1+\lambda^3)^{1/3}\) and \(m=(1-\lambda^{3/2})^{2/3}\).
Pairing the integrands at \(k\) and \(\pi-k\), write
\(a=\epsilon_{3,\lambda}(k)\),
\(b=\epsilon_{3,\lambda}(\pi-k)\), \(s=a+b\), and \(p=ab\).
The sum of the two integrands of \(\partial_x\Delta\) has a positive
denominator and numerator
\begin{equation}
  P(x)=4x^3-3sx^2+2(s^2-p)x-ps.
  \label{eq:application-n3-paired-numerator}
\end{equation}
Since \(P'(x)>0\), with discriminant $96p - 60s^2 <0$,   
$P(s/4)=3s(a-b)^2/8\geq0$, and
\(s\leq2A\), we have \(\partial_x\Delta>0\) for \(x\geq A/2\).
Moreover,
\begin{equation}
  \partial_x^2\Delta_{3,\lambda}(x)
  =\frac1\pi\int_0^\pi
  \frac{\epsilon_{3,\lambda}(k)^2+2x\epsilon_{3,\lambda}(k)-2x^2}
  {[x^2-x\epsilon_{3,\lambda}(k)+\epsilon_{3,\lambda}(k)^2]^2}
  \,\mathrm dk>0,
  \qquad 0\leq x\leq A/2,
  \label{eq:application-n3-convexity}
\end{equation}
provided \(A/m<1+\sqrt3\).  
Note that $x/\epsilon \le A/(2m) <(1+\sqrt 3)/2$, so 
this condition holds throughout
\(0<\lambda\leq4/5\), since
\((A/m)^3<945/49<(1+\sqrt3)^3\).  Thus the global minimizer is unique
on the bracket \([0.784,0.785]\).

As a check with 85-decimal-digit interval arithmetic, subdivision of \([0,\pi]\) into
1000 equal cells gives, uniformly for \(\lambda\in[0.784,0.785]\),
\begin{equation}
  \partial_x\Delta_{3,\lambda}(0.4)<-0.016,
  \qquad
  \partial_x\Delta_{3,\lambda}(0.415)>0.019.
  \label{eq:application-n3-stationary-bracket}
\end{equation}
On the resulting rectangle \(0.4\leq x\leq0.415\),
\(0.784\leq\lambda\leq0.785\), the same calculation gives
\(-1.105<\partial_\lambda\Delta_{3,\lambda}(x)<-1.000\).
The minimizing branch therefore lies in this rectangle and
\(\gamma_3'(\lambda)<-1\).  The endpoint signs quoted above are also
confirmed by interval quadrature.  Hence this bracket contains exactly one
zero of the exponential rate.

We then use this bracket and the associated stationary points to initialize
a two-variable Newton iteration for the unknowns \((x,\lambda)\), solving
\(\partial_x\Delta_{3,\lambda}(x)=0\) and
\(\Delta_{3,\lambda}(x)=0\) simultaneously.  The integrals are reevaluated
at each iteration.  The calculation used 80 decimal digits and a Newton
tolerance of \(10^{-60}\).  It gives
\begin{equation}
\begin{aligned}
  x_*&=0.4067252753367076803297\ldots,\\
  \lambda_-&=0.7846183437165991879822\ldots.
\end{aligned}
  \label{eq:application-n3-critical-root}
\end{equation}
Using the unrounded solution, we obtain
\begin{equation}
\begin{aligned}
  h_{3,\lambda_-}(x_*)&=-0.2425578657749142943447\ldots,\\
  \ln\lambda_-&=-0.2425578657749142943447\ldots.
\end{aligned}
  \label{eq:application-n3-residuals}
\end{equation}
The solution satisfies Eq.~\eqref{eq:seam-critical-saddle-conditions}
and the accompanying nondegeneracy conditions to the reported numerical
precision.

The upper value is then obtained solely by duality as
\(\lambda_+=\lambda_-^{-1}\), without a further search on
\([1,\infty)\).  Thus, for \(N=3\), the resulting reciprocal candidate critical
points are
\begin{equation}
\begin{aligned}
  \lambda_-&=0.784618343716599\ldots,\\
  \lambda_+&=\lambda_-^{-1}=1.274504997248950\ldots.
\end{aligned}
  \label{eq:prl-n3-critical-points}
\end{equation}
The lower value is the zero of the exponential rate established in
Section~\ref{sec:seam-response-asymptotics} for \(0<\lambda<1\). The upper
value is its dual counterpart.  This calculation locates the candidate
critical couplings without requiring an exact seam-response prefactor.
The finite-size results below support the physical significance of these two couplings. 
A detailed determination of the full PBC phase diagram is deferred to future work.

\begingroup
\setlength{\intextsep}{6pt}
\enlargethispage{\baselineskip}
\begin{figure}[H]
 \centering
 \includegraphics[width=0.92\textwidth]{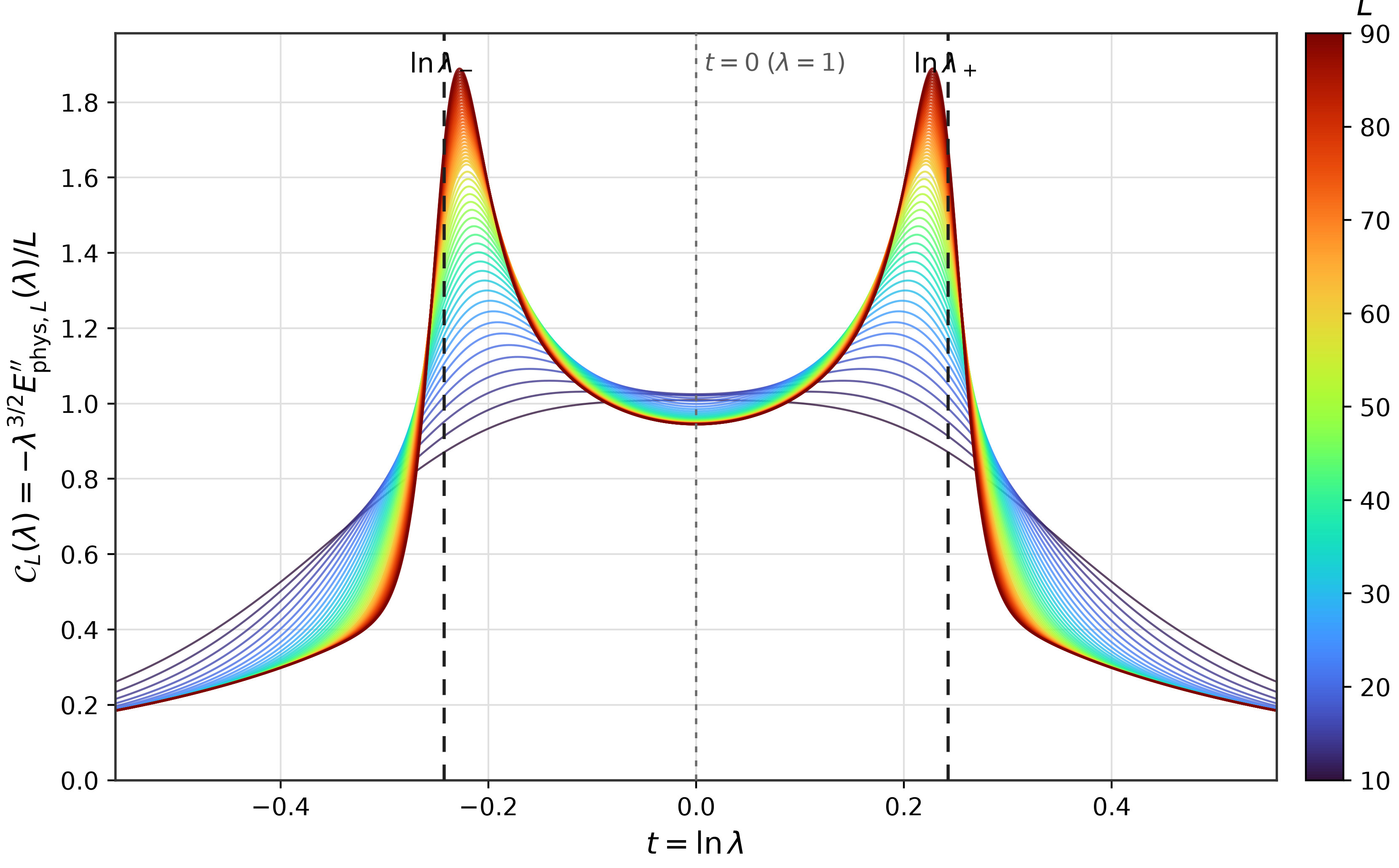}
 \caption{\label{fig:prl-n3-curvature}
Duality-normalized curvature
\(c_L(\lambda)=-\lambda^{3/2}E_{\mathrm{phys},L}''(\lambda)/L\)
of the ground-state energy of \(H_{\mathrm{phys}}=-H_{\mathrm{PBC}}\)
for the homogeneous three-state chain and even integers \(L\in[10,90]\).
The dashed lines at \(t_\pm=\ln\lambda_\pm\) mark the candidate 
critical couplings. Energies for \(\lambda>1\) are obtained
from those at \(\lambda^{-1}\) using the finite-size duality.}
\end{figure}
\endgroup

For the energy data in Fig.~\ref{fig:prl-n3-curvature}, we solved the
finite spectral equations in the zero-charge sector by damped Newton
iteration with an analytic Jacobian.
To improve numerical stability, we used Chebyshev nodes for avoiding Runge's phenomenon in polynomial interpolation.
These nodes were linearly mapped from \([-1,1]\) to \([0.1,0.9]\),
and their positive real cube roots were used as the \(L-1\) spectral
points at which the closure identity in
Eq.~\eqref{eq:matching-polynomial-closure} was imposed.
The energies at \(\lambda=1\) agreed to at least 70 decimal places
when computed at different working precisions.
For each \(L\), a previously computed coefficient vector at
\(\lambda=1\) was refined and continued towards smaller couplings,
using a secant predictor and Newton correction.
For \(\lambda>1\), we solved at the inverse coupling and used the finite-size duality of Lemma~\ref{lem:finite-duality} to obtain the energy, without interpolation. 
The second derivative was evaluated by centered finite differences in
\(\lambda\) with step \(0.001\).

\begin{figure}[H]
 \centering
 \includegraphics[width=0.92\textwidth]{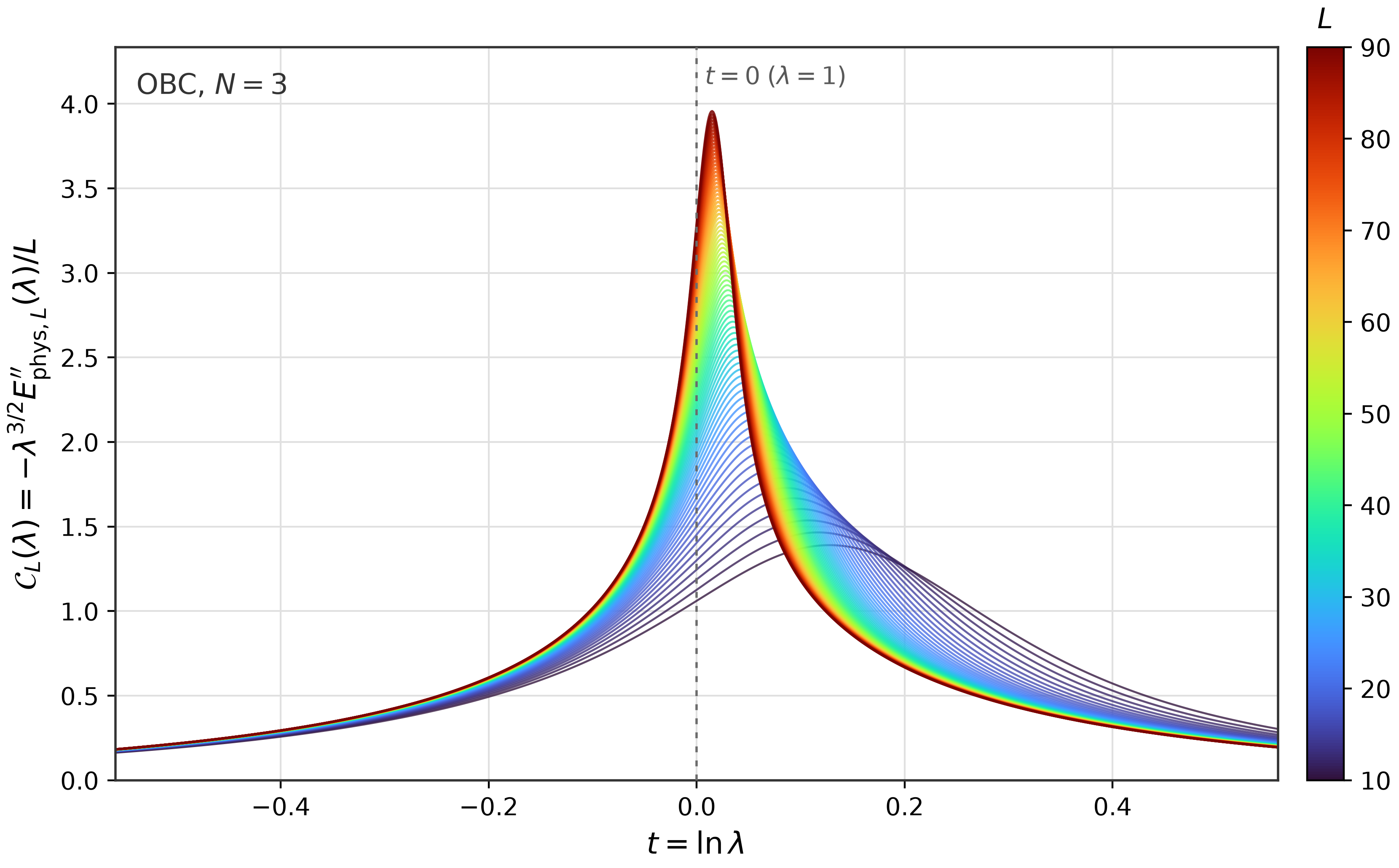}
 \caption{\label{fig:obc-n3-curvature}
 Curvature of the ground-state energy of \(-H_{\mathrm{OBC}}\) for the
 homogeneous \(N=3\) chain, with \(L=10,11,\ldots,90\), in the same
 coordinates as Fig.~\ref{fig:prl-n3-curvature}.  The second derivative is
 evaluated by centered finite differences in \(\lambda\) with step
 \(0.001\). The dashed line marks \(\lambda=1\).}
\end{figure}

Figure~\ref{fig:prl-n3-curvature} compares the finite-\(L\) second derivative
of the ground-state energy with these thermodynamic-limit predictions.  For
small \(L\), the curvature retains a single unresolved central feature. The
splitting into two off-center peaks becomes visible only as \(L\) increases.
For comparison, Fig.~\ref{fig:obc-n3-curvature} shows the OBC curvature
using the same coordinate transformation and coupling step.  It retains a
single peak that sharpens and approaches \(\lambda=1\) as \(L\) increases,
consistent with the exact OBC quasienergy gap closing on the positive-real
coupling axis only at \(\lambda=1\)~\cite{AlcarazBatchelor2018}.

\section{Conclusion and discussion}
\label{sec:conclusion-discussion}

Periodic boundary conditions change the algebraic form rather than eliminating the exact
solvability of the Baxter-Fendley chain.  Under OBC, the spectrum is assembled
from independent free-parafermion quasienergies.  Under PBC, the current path
becomes the matching cycle \(C_{2L}\), while the root-of-unity spectral orbit
gives the second cycle \(C_N\), with the independent open-chain occupation rule 
thereby replaced by a global closure condition.  We constructed the
corresponding operator-valued matching polynomial and, after site-local
pairing, identified it with a cyclic Yang-Baxter transfer matrix.  Its first
nonconstant coefficient is the Hamiltonian, and the RTT relation implies that
all of its coefficients commute. They are therefore conserved by the
Hamiltonian.

On the chart \(\lambda_{2j-1}\neq0\), \(j=1,\ldots,L\), every physical 
transfer eigenvalue in a fixed charge sector determines a common zero of the 
\(L-1\) polynomial equations in Eq.~\eqref{eq:finite-spectral-equations}. 
Each equation has degree \(N\). Their projective homogenizations have no 
common zero at infinity, and Bézout’s theorem gives \(N^{L-1}=\dim{\cal H}_q\) 
as the total intersection multiplicity of the finite zeros. 
Proposition 6.4 further identifies the normalized energy resultant of this system 
with the charge-sector characteristic polynomial, so the finite spectral equations 
reproduce the complete energy spectrum with algebraic multiplicity. 
Although the BBS fusion derivation is initially made on the nonzero odd-coupling chart, 
this resultant identity extends polynomially to arbitrary physical couplings. 
This finite-coefficient formulation complements the SoV spectral equations in
Appendix~\ref{app:separation-of-variables-solution} and, for odd \(N\) on the
generic simple-root chart, the ODBA equations in
Appendix~\ref{app:odba-finite-size-verification}. 
It also provides fixed-degree equations suitable for elimination and
continuation~\cite{vonGehlenEtAl2006,XuEtAl2015}.

For positive reflection-symmetric couplings, the Perron-Frobenius argument
identifies a unique algebraically simple maximal-real-part eigenvalue in the
neutral charge sector and proves its analytic continuation from the all-zero
OBC ground state to the homogeneous PBC endpoint.  
The resultant identity together with the Perron–Frobenius simplicity theorem implies that the coefficient Jacobian is nonsingular along the physical seam branch. The implicit-function equation and its discretization in Section~\ref{sec:physical-branch-continuation} therefore follow this branch without enumerating the remaining solutions of the finite spectral system.

The exact finite-size seam-response formula determines the exponential rate
for \(0<\lambda<1\), as shown in
Section~\ref{sec:seam-response-asymptotics}, and yields the saddle equations
used for the homogeneous \(N=3\) chain.  Their high-precision numerical solution, the duality relation, and the finite-size ground-state curvature support two reciprocal candidate PBC critical couplings
\(\lambda_-=0.784618343716599\ldots\) and
\(\lambda_+=1.274504997248950\ldots\).  The curvature evolves from one
unresolved central feature at small \(L\) to two off-center peaks as \(L\)
increases.  This differs from OBC, for which the positive-real quasienergy gap
closes only at \(\lambda=1\)~\cite{AlcarazBatchelor2018}.
The detailed excitation spectrum, critical scaling and full periodic phase diagram will be investigated in future work.

The matching-polynomial and transfer-matrix formulation provides a starting
point for further algebraic and asymptotic investigations of the periodic
chain, including excited-state branches, correlation functions,
thermodynamic limits, and finite-size structure for general \(N\).

\backmatter


\bmhead{Acknowledgements}

This work was supported by Australian Research Council Grant DP240100838.


\vskip 10mm

\begin{appendices}

\section{\texorpdfstring{\enspace}{}Off-diagonal Bethe ansatz and finite-size verification}
\label{app:odba-finite-size-verification}
\setcounter{equation}{0}
\renewcommand{\theequation}{A\arabic{equation}}
\renewcommand{\theHequation}{appendix.A.\arabic{equation}}
\setcounter{table}{0}
\renewcommand{\thetable}{A\arabic{table}}
\renewcommand{\theHtable}{appendix.A.\arabic{table}}

For odd \(N\), the periodic \(\tau^{(2)}\) ODBA of Xu et
al.~\cite{XuEtAl2015} applies to the present clock specialization.  The
construction is useful here for two reasons.  First, it provides an independent
\(T\)--\(Q\) parametrization of the transfer eigenvalues encoded by the
root-of-unity closure in the main text.  Second, it permits a direct finite-size
test of the identification of the periodic Baxter-Fendley chain with the
cyclic BBS transfer matrix.  
The matching-polynomial closure, Bézout count, and resultant completeness theorem are derived independently 
in the main text. The ODBA
parametrization is not used in those derivations.  Related results on
inhomogeneous \(T\)--\(Q\) relations in finite-dimensional integrable systems
provide additional context~\cite{CaoEtAl2015}.

\subsection{Cyclic BBS specialization}

We use the cyclic BBS \(L\)-operator in the form
\begin{equation}
  L_j^{\mathrm{BBS}}(x)=
  \begin{pmatrix}
    A_j(x)&B_j(x)\\
    C_j(x)&D_j(x)
  \end{pmatrix},
  \label{eq:app-a-bbs-l-operator}
\end{equation}
with
\begin{align*}
  A_j(x)&=1+x\varkappa_j\mathsf v_j,
  &B_j(x)&=x\mathsf u_j^{-1}(a_j-b_j\mathsf v_j),\\
  C_j(x)&=\mathsf u_j(c_j-d_j\mathsf v_j),
  &D_j(x)&=xa_jc_j+\varkappa_j^{-1}\mathsf v_jb_jd_j.
\end{align*}
For nonzero \(\varkappa_j\), the local matrix \(L_{a,j}(u)\) used in the main
text is obtained from the exact parameter choice
\begin{equation}
  \begin{gathered}
    x=u,\qquad
    \mathsf u_j=\sigma_j,\qquad
    \mathsf v_j=\tau_j,\\
    \varkappa_j=-\lambda_{2j-1},\qquad
    a_j=-\lambda_{2j},\qquad
    b_j=d_j=0,\qquad
    c_j=1.
  \end{gathered}
  \label{eq:app-a-bbs-specialization}
\end{equation}
Thus the monodromy and the charge-sector transfer eigenvalue
\(\Lambda_q(u)\) are those of
Eqs.~\eqref{eq:monodromy-transfer-matrix}
and~\eqref{eq:transfer-eigenvalue}, respectively, while the quantum
determinant is defined in Eq.~\eqref{eq:quantum-determinant-definition},
evaluated for the monodromy in
Eq.~\eqref{eq:monodromy-quantum-determinant}, and restricted to
\(\mathcal H_q\) in Eq.~\eqref{eq:sector-quantum-determinant}.
We remain in this subsection on the generic nonzero-\(\varkappa\) chart. 
At singular coupling values, we use instead the polynomial transfer-matrix
closure and the finite spectral equations of the main text.

Let \(\tau_q^{(j)}(u)\) denote the fused-transfer eigenvalues, normalized by
\(\tau_q^{(0)}=0\), \(\tau_q^{(1)}=1\), and
\(\tau_q^{(2)}=\Lambda_q\). The continuant form of the fusion hierarchy
identifies its terminal combination with the spectral-cycle matching
polynomial. The cyclic-BBS root-of-unity truncation theorem gives
\begin{equation}
  \Phi_q[\Lambda_q](u)
  =\tau_q^{(N+1)}(u)
   -\alpha_{q,0}(u)\tau_q^{(N-1)}(\omega u)
  =\operatorname{tr}_a\prod_{j=1}^{L}\mathcal L_{a,j}(u^N)
  =\mathcal Z_{\boldsymbol\lambda}^{(N)}(u^N).
  \label{eq:app-a-root-of-unity-closure}
\end{equation}
This is precisely the closure relation obtained in the main text from the
matching-polynomial construction.

\subsection{Inhomogeneous \texorpdfstring{\(T\)--\(Q\)}{T-Q} relation}

Choose an auxiliary factorization parameter \(s\in\mathbb C^\times\) and write
\begin{equation}
  a_s(u)=su^L,\qquad
  d_{q,s}(u)=\frac{\omega^q\lambda_{\mathrm{tot}}}{s}u^L,
  \qquad
  a_s(\omega^ru)d_{q,s}(\omega^{r-1}u)=\alpha_{q,r}(u).
  \label{eq:app-a-baxter-factorization}
\end{equation}
Set \(\Lambda_{q,r}=\Lambda_q(\omega^ru)\),
\(a_{s,r}=a_s(\omega^ru)\), and
\(d_{q,s,r}=d_{q,s}(\omega^ru)\). For cyclic indices
\(r,r'\in\mathbb Z_N\), define
\begin{subequations}
\begin{equation}
  (D_{q,s})_{rr'}
  =\Lambda_{q,r}\delta_{rr'}
   -a_{s,r}\delta_{r',r-1}
   -d_{q,s,r}\delta_{r',r+1}.
  \label{eq:app-a-cyclic-baxter-matrix}
\end{equation}
The nonwinding determinant terms give \(\Phi_q[\Lambda_q]\), while the two
oriented windings are explicit. Using Eq.~\eqref{eq:app-a-root-of-unity-closure},
one obtains
\begin{equation}
  \det D_{q,s}(u)=:\mathcal F_s(u)
  =\mathcal Z_{\boldsymbol\lambda}^{(N)}(u^N)
   -(-1)^{L(N-1)}s^Nu^{LN}
   -(-1)^{L(N-1)}\lambda_{\mathrm{tot}}^Ns^{-N}u^{LN}.
  \label{eq:app-a-baxter-source}
\end{equation}
\end{subequations}
Let \(\mathbf1=(1,\ldots,1)^T\) and define
\[
 \mathbf Q(u)=\operatorname{adj}D_{q,s}(u)\,\mathbf1,
 \qquad Q(u)=\mathbf e_0^{\,T}\mathbf Q(u).
\]
The adjugate identity gives
\(D_{q,s}(u)\mathbf Q(u)=F_s(u)\mathbf1\).
If \((Pz)_r=z_{r+1}\), then
\(D_{q,s}(\omega u)=PD_{q,s}(u)P^{-1}\), and hence
\(Q_r(u)=Q(\omega^ru)\). Since \(D_{q,s}(0)=I\), we also have
\(Q(0)=1\). The zeroth row is the inhomogeneous \(T\)--\(Q\) relation 
\begin{equation}
  \Lambda_q(u)Q(u)
  =su^LQ(\omega^{-1}u)
   +\frac{\omega^q\lambda_{\mathrm{tot}}}{s}u^LQ(\omega u)
   +\mathcal F_s(u), 
  \label{eq:app-a-inhomogeneous-tq}
\end{equation}
and \(\deg Q\le(N-1)L\).
On the generic full-degree simple-root chart one may write
\[
  Q(u)=\prod_{j=1}^{M}\left(1-\frac{u}{u_j}\right),
  \qquad M=(N-1)L.
\]
Evaluation of Eq.~\eqref{eq:app-a-inhomogeneous-tq} at the roots \(u_j\) gives
\begin{equation}
  0=su_j^LQ(\omega^{-1}u_j)
   +\frac{\omega^q\lambda_{\mathrm{tot}}}{s}u_j^LQ(\omega u_j)
   +\mathcal F_s(u_j),
  \qquad j=1,\ldots,M.
  \label{eq:app-a-bethe-equations}
\end{equation}

\subsection{Leading-coefficient selection and energy}

The Bethe equations \eqref{eq:app-a-bethe-equations} must be supplemented by
the condition that the reconstructed transfer eigenvalue has the fixed
leading coefficient required by the two perfect matchings of the current
cycle. Put
\[
  f_s=[u^{LN}]\mathcal F_s(u),
\]
and define
\[
  \Delta_{q,s}
  =(-1)^L(\omega^q\lambda_{\mathrm o}+\lambda_{\mathrm e})
   -s\omega^L-\frac{\omega^{q-L}\lambda_{\mathrm{tot}}}{s},
\]
\[
  f_s=(-1)^L(\lambda_{\mathrm o}^N+\lambda_{\mathrm e}^N)
   -(-1)^{L(N-1)}
    \left(s^N+\lambda_{\mathrm{tot}}^Ns^{-N}\right).
\]
When \(f_s\neq0\), the leading-coefficient condition is
\begin{equation}
  \prod_{j=1}^{M}u_j=(-1)^M\frac{\Delta_{q,s}}{f_s}.
  \label{eq:app-a-leading-coefficient-condition}
\end{equation}
It is useful to regard Eq.~\eqref{eq:app-a-leading-coefficient-condition} as
an \emph{asymptotic/sector selection condition}, rather than as an independent
local Bethe equation. It enforces the prescribed highest coefficient of
\(\Lambda_q(u)\) and hence the chosen charge-sector branch.

Every physical solution represented on this generic chart has energy
\begin{equation}
  E_q=-\sum_{j=1}^{M}u_j^{-1}.
  \label{eq:app-a-energy}
\end{equation}

\subsection{Finite-size verification}

We have checked the specialized ODBA equations against direct finite-size diagonalization
for generic inhomogeneous couplings. In each test
the states were resolved by the global \(Z_N\) charge. For every state we
reconstructed the polynomial \(Q(u)\) of degree \(M=(N-1)L\), verified the
full polynomial identity \eqref{eq:app-a-inhomogeneous-tq}, checked the
root-product condition \eqref{eq:app-a-leading-coefficient-condition}, and
reconstructed the energy from Eq.~\eqref{eq:app-a-energy}.

A representative set of checks used
\[
  (\lambda_1,\ldots,\lambda_6)=(1.1,0.7,0.9,1.2,1.05,0.8)
\]
for \(L=3\) (with the corresponding truncations for \(L=2\)), and a generic
complex factorization parameter \(s=1.371+0.217\,i\).  For \(N=3\), all exact diagonalization (ED) 
eigenstates were checked for \(L=2,3,4\).  As a nontrivial test beyond the
three-state chain, all ED eigenstates of the \(N=5\) model were checked for
\(L=2\) and \(L=3\). The latter comprises \(5^3=125\) states, or 25 states in
each charge sector.  Here \(E_{\mathrm{ODBA}}\) denotes the energy reconstructed
from Eq.~\eqref{eq:app-a-energy}, while \(E_{\mathrm{ED}}\) is the corresponding
charge-resolved eigenvalue obtained by exact diagonalization.

\begin{table}[htbp]
\caption{Charge-resolved finite-size verification of the specialized ODBA
equations.  The reported maxima are taken over all checked states.}
\label{tab:app-a-odba-finite-size-verification}
\centering
\small
\begin{tabular}{ccrcc}
\hline
\(N\)&\(L\)&states checked&max. polynomial \(T\)--\(Q\) residual
&max. \(\lvert E_{\mathrm{ODBA}}-E_{\mathrm{ED}}\rvert\)\\
\hline
3&2&9&\(7.7\times10^{-15}\)&\(4.9\times10^{-15}\)\\
3&3&27&\(6.6\times10^{-14}\)&\(2.4\times10^{-14}\)\\
3&4&81&\(3.0\times10^{-13}\)&\(4.5\times10^{-14}\)\\
5&2&25&\(4.7\times10^{-14}\)&\(1.3\times10^{-14}\)\\
5&3&125&\(3.7\times10^{-13}\)&\(9.4\times10^{-14}\)\\
\hline
\end{tabular}
\end{table}

The root-product condition \eqref{eq:app-a-leading-coefficient-condition} is
satisfied at the same numerical level (typically \(10^{-14}\) or better.
For the \(N=5\), \(L=3\) calculation the worst discrepancy was below
\(5\times10^{-15}\)). Direct evaluation of the Bethe equations
\eqref{eq:app-a-bethe-equations} at the degree-12 \(Q\)-roots for \(N=5\),
\(L=3\) gives relative residuals of order \(10^{-13}\). The somewhat larger
residuals in the highest-degree tests arise from conditioning of polynomial
coefficient and root calculations and do not indicate a systematic
discrepancy.

We also varied the auxiliary parameter \(s\) in the \(N=5\), \(L=2\)
calculation, including real and complex values, and found the same physical
spectrum while the intermediate \(Q\)-polynomial changed as expected. This
confirms numerically that \(s\) is a factorization parameter and not a
physical coupling.

These checks are not used as a proof of completeness.  Their role is to
verify, on nontrivial finite systems and in every charge sector, the
specialized ODBA formulas 
\eqref{eq:app-a-bbs-specialization}--\eqref{eq:app-a-energy} 
and their agreement with the matching-polynomial
spectral equations derived independently in the main text.

\subsection{Scope of the ODBA formulation}

The polynomial ODBA description above is stated for odd \(N\) on the generic
nonzero-\(\varkappa\) and simple-root chart~\cite{XuEtAl2015}.  Complete
\(N\)-strings, vanishing \(\varkappa_j\), or other singular Bethe-root
configurations can require a different local parametrization of \(Q\).  These
limitations concern the local Bethe-root parametrization, whereas the
transfer-matrix closure and the finite spectral equations remain polynomial in
the physical couplings.  Appendix A is therefore used as a Bethe-root
parametrization and an independent finite-size check of the transfer-matrix
formulation.

\section{\texorpdfstring{\enspace}{}Separation-of-variables formulation}
\label{app:separation-of-variables-solution}
\setcounter{equation}{0}
\renewcommand{\theequation}{B\arabic{equation}}
\renewcommand{\theHequation}{appendix.B.\arabic{equation}}

The cyclic-BBS SoV construction supplies a
second independent formulation~\cite{vonGehlenEtAl2006}.
For the SoV representation below, we restrict to nonzero couplings
for which the averaged upper-right entry of every partial monodromy,
after division by \(u^N\), has only nonzero simple zeros. These restrictions
concern only this representation and do not alter the polynomial closure
established in the main text~\cite{vonGehlenEtAl2006}.
For the monodromy in
Eq.~\eqref{eq:monodromy-transfer-matrix}, direct multiplication gives
\begin{equation}
B(u)=-u\lambda_{2L}G_{L-1}(u)\sigma_L^\dagger,
\qquad
G_{L-1}(u)=
\sum_{F\in{\cal M}(P_{2L-1})}
(-u)^{|F|}\prod_{e_n\in F}h_n .
\end{equation}
Here \(P_{2L-1}\) has vertices \(1,\ldots,2L-1\) and edges \(e_1,\ldots,e_{2L-2}\), with \(e_n\) carrying the current weight \(h_n\).
Since $[G_{L-1}(u),\sigma_L]=0$, fixing the endpoint clock value
reduces the $B$-eigenvalue problem to the fixed-boundary open
Baxter-Fendley chain.  If
\begin{equation}
\prod_{s=0}^{N-1}G_{L-1}(\omega^su)
=\prod_{k=1}^{L-1}(1-r_k^Nu^N)\,\mathbf 1 ,
\end{equation}
the $B$-eigenvalues may be written
\begin{equation}
B(u)|\rho_0,\boldsymbol{\rho}\rangle_B
=-u\lambda_{2L}\omega^{-\rho_0}
\prod_{k=1}^{L-1}(1-ur_k\omega^{\rho_k})
|\rho_0,\boldsymbol{\rho}\rangle_B ,
\end{equation}
with $\rho_0,\rho_k\in\mathbb Z_N$.  For homogeneous
$\lambda_{2j-1}=1$, $\lambda_{2j}=\lambda$,
\begin{equation}
r_k^N=
1+\lambda^N+2\lambda^{N/2}\cos\frac{k\pi}{L},
\qquad k=1,\ldots,L-1,
\end{equation}
so the separated amplitudes have precisely the open-chain free-parafermion dispersion, 
evaluated at the fixed-boundary momenta \(k\pi/L\).

After Fourier transforming with respect to $\rho_0$ and projecting onto
charge sector $q$, the coefficient of the SoV state factorizes after
including the standard Sklyanin measure~\cite{vonGehlenEtAl2006}:
\begin{equation}
\widetilde Q(\rho_1,\ldots,\rho_{L-1})
=\prod_{k=1}^{L-1}Q_k(\rho_k).
  \label{eq:app-b-sov-factorization}
\end{equation}
Evaluating $t(u)|\Psi_q\rangle=\Lambda_q(u)|\Psi_q\rangle$ at
$u_{k,s}=\omega^{-s}/r_k$, a zero of the corresponding $B$-eigenvalue,
the RTT algebra implies that $A$ and $D$ shift the $k$th separated
coordinate in opposite directions.  One obtains the cyclic Baxter equations
\begin{equation}
\Lambda_q(u_{k,s})Q_k(s)
=\Delta_{k,+}(s)Q_k(s+1)
+\Delta_{k,-}(s)Q_k(s-1),
\qquad s\in\mathbb Z_N ,
  \label{eq:app-b-cyclic-baxter-equations}
\end{equation}
We fix the separated-state gauge explicitly. Write the averaged
monodromy as
\[
 \mathcal T(z)=\prod_{j=1}^L\mathcal L_j(z)
 =\begin{pmatrix}\mathcal A(z)&\mathcal B(z)\\
                  \mathcal C(z)&\mathcal D(z)\end{pmatrix},
 \qquad g_j=\lambda_{2j-1}\lambda_{2j},
\]
where \(\mathcal L_j\) is the scalar matrix in \eqref{eq:classical-clock-lax-operator}. Put
\(z_k=r_k^{-N}\) and choose an \(N\)th root \(\chi_k\) satisfying
\[
 \chi_k^N=(-1)^{(N-1)(L+1)}
 \frac{\mathcal A(z_k)r_k^{N(L+1)}}{g_L^N}.
\]
The amplitudes in (B6) may then be taken as
\[
 \Delta_{k,+}(s)=\frac{\omega^q}{\chi_k}
      \left(\prod_{j=1}^{L-1}g_j\right)u_{k,s}^{L-1},
 \qquad
 \Delta_{k,-}(s)=\omega\chi_kg_Lu_{k,s}^{L+1}.
\]
They obey both the adjacent-product relation
\[
 \Delta_{k,+}(s)\Delta_{k,-}(s+1)
 =\omega^{q-L}\lambda_{\mathrm{tot}}u_{k,s}^{2L}
\]
and the two winding relations
\[
 \prod_{s=0}^{N-1}\Delta_{k,+}(s)=\mathcal D(z_k),
 \qquad
 \prod_{s=0}^{N-1}\Delta_{k,-}(s)=\mathcal A(z_k).
\]
At \(\mathcal B(z_k)=0\), the averaged determinant gives
\(\mathcal A(z_k)\mathcal D(z_k)=
\lambda_{\mathrm{tot}}^Nz_k^{2L}\). Thus all quantities used in
this normalization are nonzero on the stated chart. The choice
\(\chi_k\mapsto\omega^m\chi_k\) is compensated by the periodic
gauge transformation \(Q_k(s)\mapsto\omega^{ms}Q_k(s)\).

These formulas specialize Eqs. (102)--(104) of~\cite{vonGehlenEtAl2006}. The dictionary is
\(n=L\), \(\lambda_{\mathrm{[18]}}=u\),
\(\rho_{\mathrm{[18]}}=q\),
\(r_{L,k}^{\mathrm{[18]}}=-r_k^{-1}\), and
\(\rho_{L,k}^{\mathrm{[18]}}=s\).
The zeroth-coordinate relabelling is
\(\rho_{L,0}^{\mathrm{[18]}}=\rho_0-\sum_k\rho_k\), with
\(r_{L,0}^{\mathrm{[18]}}=(-1)^L\lambda_{2L}\prod_k r_k\).
The corresponding one-coordinate phases after the charge Fourier
transform are absorbed into the separated factors \(Q_k\).

Equivalently, Eq.~\eqref{eq:app-b-cyclic-baxter-equations} is the
$N\times N$ cyclic homogeneous system
\begin{equation}
{\cal D}^{(k)}_q[\Lambda_q]\,{\bf Q}_k=0,
  \label{eq:app-b-sov-linear-system}
\end{equation}
where ${\bf Q}_k=(Q_k(0),\ldots,Q_k(N-1))^{\mathsf T}$ and
\begin{equation}
\bigl({\cal D}^{(k)}_q\bigr)_{ss'}
=\Lambda_q(u_{k,s})\delta_{ss'}
-\Delta_{k,+}(s)\delta_{s',s+1}
-\Delta_{k,-}(s)\delta_{s',s-1}.
  \label{eq:app-b-sov-matrix-elements}
\end{equation}
All indices are understood modulo $N$.  A nonzero separated wavefunction
exists precisely when
\begin{equation}
\det {\cal D}^{(k)}_q[\Lambda_q]=0,
\qquad k=1,\ldots,L-1.
  \label{eq:app-b-sov-solvability}
\end{equation}
The cyclic-BBS SoV theorem~\cite{vonGehlenEtAl2006} identifies
simultaneous solvability of the $L-1$
systems in Eq.~\eqref{eq:app-b-cyclic-baxter-equations} with the fusion closure
in Eq.~\eqref{eq:matching-polynomial-closure}.
Thus the OBC free-parafermion
modes survive as PBC separated variables, while their independent
occupations are replaced by the global root-of-unity constraint.

\end{appendices}


\begin{thebibliography}{99}
\addcontentsline{toc}{section}{\refname}

\bibitem{Alcaraz2026Nonhomogeneous}
F. C. Alcaraz,
``Free-fermionic and free-parafermionic multispin quantum chains with
nonhomogeneous interacting ranges,''
\emph{Phys. Rev. E} \textbf{114} (2026), 014146,
\url{https://doi.org/10.1103/zxcm-zbbb}.

\bibitem{AlcarazBatchelor2018}
F. C. Alcaraz and M. T. Batchelor,
``Anomalous bulk behaviour in the free parafermion \(Z(N)\) spin chain,''
\emph{Phys. Rev. E} \textbf{97} (2018), 062118,
\url{https://doi.org/10.1103/PhysRevE.97.062118}.

\bibitem{AlcarazBatchelorLiu2017}
F. C. Alcaraz, M. T. Batchelor, and Z.-Z. Liu,
``Energy spectrum and critical exponents of the free parafermion
\(Z_N\) spin chain,''
\emph{J. Phys. A: Math. Theor.} \textbf{50} (2017), 16LT03,
\url{https://doi.org/10.1088/1751-8121/aa645a}.

\bibitem{BatchelorHenryLu2023}
M. T. Batchelor, R. A. Henry, and X. Lu,
``A brief history of free parafermions,''
\emph{AAPPS Bull.} \textbf{33} (2023), 29,
\url{https://doi.org/10.1007/s43673-023-00105-3}.

\bibitem{Baxter1989}
R. J. Baxter,
{A simple solvable \(Z_N\) Hamiltonian}, \emph{Phys. Lett. A}
\textbf{140} (1989), 155--157,
\url{https://doi.org/10.1016/0375-9601(89)90884-0}.

\bibitem{Baxter2004}
R. J. Baxter,
``Transfer matrix functional relations for the generalized
\(\tau_2(t_q)\) model,'' \emph{J. Stat. Phys.} \textbf{117} (2004),
1--25,
\url{https://doi.org/10.1023/B:JOSS.0000044062.64287.b9}.


\bibitem{BazhanovStroganov1990}
V. V. Bazhanov and Yu. G. Stroganov,
``Chiral Potts model as a descendant of the six-vertex model,''
\emph{J. Stat. Phys.} \textbf{59} (1990), 799--817,
\url{https://doi.org/10.1007/BF01025851}.


\bibitem{CaoEtAl2015}
J. Cao, W.-L. Yang, K. Shi, and Y. Wang,
``On the complete-spectrum characterization of quantum integrable spin chains
via the inhomogeneous \(T\)--\(Q\) relation,''
\emph{J. Phys. A: Math. Theor.} \textbf{48} (2015), 444001,
\url{https://doi.org/10.1088/1751-8113/48/44/444001}.


\bibitem{CobaneraOrtizNussinov2011}
E. Cobanera, G. Ortiz, and Z. Nussinov, 
``The bond-algebraic approach to dualities,''
\emph{Advances in Physics} \textbf{60} (2011), 679–798,
\url{https://doi.org/10.1080/00018732.2011.619814}.


\bibitem{CLO2005}
D. A. Cox, J. Little, and D. O'Shea,
\emph{Using Algebraic Geometry}, 2nd ed., Graduate Texts in Mathematics,
vol.~185, Springer, New York (2005).

\bibitem{ElmanChapmanFlammia2021}
S. J. Elman, A. Chapman, and S. T. Flammia,
``Free fermions behind the disguise,'' \emph{Commun. Math. Phys.}
\textbf{388} (2021), 969--1003,
\url{https://doi.org/10.1007/s00220-021-04220-w}.


\bibitem{EvansHoeghKrohn1978}
D. E. Evans and R. H{\o}egh-Krohn,
``Spectral properties of positive maps on \(C^*\)-algebras,''
\emph{J. Lond. Math. Soc.} (2) \textbf{17} (1978), 345--355,
\url{https://doi.org/10.1112/jlms/s2-17.2.345}.


\bibitem{EvansKawahigashi1998}
D. E. Evans and Y. Kawahigashi,
\emph{Quantum Symmetries on Operator Algebras}, Oxford Mathematical
Monographs, Clarendon Press, Oxford (1998),
\url{https://doi.org/10.1093/oso/9780198511755.001.0001}.

\bibitem{FacchiniLeroy2015}
A. Facchini and A. Leroy,
``Leapfrog constructions: From continuant polynomials to permanents of
matrices,'' \emph{Electron. J. Combin.} \textbf{22} (2015), Paper P1.39,
\url{https://doi.org/10.37236/4637}.

\bibitem{Fendley2014}
P. Fendley,
{Free parafermions}, \emph{J. Phys. A: Math. Theor.}
\textbf{47} (2014), 075001,
\url{https://doi.org/10.1088/1751-8113/47/7/075001}.


\bibitem{FradkinSusskind1978}
E. Fradkin and L. Susskind,
``Order and disorder in gauge systems and magnets,''
\emph{Phys. Rev. D} \textbf{17} (1978), 2637–2658,
\url{https://doi.org/10.1103/PhysRevD.17.2637}.


\bibitem{FukaiPozsgayVona2026}
K. Fukai, B. Pozsgay, and I. Vona,
``Solving models with generalized free fermions II: Path-product expansion
and conserved charges,'' arXiv:2605.31453v2 (2026),
\url{https://arxiv.org/abs/2605.31453v2}.


\bibitem{vonGehlenEtAl2006}
G. von Gehlen, N. Iorgov, S. Pakuliak, and V. Shadura,
{Baxter-Bazhanov-Stroganov model: Separation of variables and Baxter
equation}, \emph{J. Phys. A: Math. Gen.} \textbf{39} (2006), 7257--7282,
\url{https://doi.org/10.1088/0305-4470/39/23/006}.


\bibitem{Hartshorne1977}
R. Hartshorne,
\emph{Algebraic Geometry}, Graduate Texts in Mathematics, vol.~52,
Springer, New York (1977),
\url{https://doi.org/10.1007/978-1-4757-3849-0}.


\bibitem{HeilmannLieb1972}
O. J. Heilmann and E. H. Lieb,
``Theory of monomer--dimer systems,'' \emph{Commun. Math. Phys.}
\textbf{25} (1972), 190--232,
\url{https://doi.org/10.1007/BF01877590}.


\bibitem{HenryBatchelor2023}
R. A. Henry and M. T. Batchelor,
``Exceptional points in the Baxter-Fendley free parafermion model,''
\emph{SciPost Phys.} \textbf{15} (2023), 016,
\url{https://doi.org/10.21468/SciPostPhys.15.1.016}.



\bibitem{IzerginKorepin1981}
A. G. Izergin and V. E. Korepin,
``A lattice model related to the nonlinear Schr\"odinger equation,''
\emph{Dokl. Akad. Nauk SSSR} \textbf{259} (1981), 76--79.


\bibitem{Kato1995}
T. Kato,
\emph{Perturbation Theory for Linear Operators}, 2nd ed., Classics in
Mathematics, Springer, Berlin and Heidelberg (1995),
\url{https://doi.org/10.1007/978-3-642-66282-9}.




\bibitem{MannElmanWoodChapman2025}
R. L. Mann, S. J. Elman, D. R. Wood, and A. Chapman,
``A graph-theoretic framework for free-parafermion solvability,''
\emph{Proc. R. Soc. A} \textbf{481} (2025), 20240671,
\url{https://doi.org/10.1098/rspa.2024.0671}.


\bibitem{RadchenkoRodriguezVillegas2021}
D. Radchenko and F. Rodriguez Villegas,
``Independence polynomials and hypergeometric series,''
\emph{Bull. Lond. Math. Soc.} \textbf{53} (2021), 1834--1848,
\url{https://doi.org/10.1112/blms.12545}.


\bibitem{Tarasov1992}
V. O. Tarasov,
``Cyclic monodromy matrices for the \(R\)-matrix of the six-vertex model and
the chiral Potts model with fixed spin boundary conditions,''
\emph{Int. J. Mod. Phys. A} \textbf{7} (1992), Suppl. 1B, 963--975,
\url{https://doi.org/10.1142/S0217751X92004129}.


\bibitem{XuEtAl2015}
X. Xu, J. Cao, S. Cui, W.-L. Yang, K. Shi, and Y. Wang,
``Off-diagonal Bethe Ansatz solution of the \(\tau_2\)-model,''
\emph{J. High Energy Phys.} \textbf{09} (2015), 212,
\url{https://doi.org/10.1007/JHEP09(2015)212}.



\end{thebibliography}
\end{document}